\documentclass[fleqn,usenatbib,twocolumn]{mnras}

\usepackage{newtxtext,newtxmath}

\usepackage[T1]{fontenc}
\usepackage{ae,aecompl}

\usepackage{xcolor}
\usepackage{pifont}

\usepackage{amsmath}
\usepackage{etoolbox}
\makeatletter
\patchcmd\@combinedblfloats{\box\@outputbox}{\unvbox\@outputbox}{}{%
}%
\makeatother

\usepackage{graphicx}	
\usepackage{eso-pic}
\usepackage{orcidlink}
\usepackage{float}
\usepackage{subcaption}
\usepackage{verbatim}
\usepackage{todonotes}

\begin{document}

\title
[How massive and clumpy are quasar winds?]
{
How massive and clumpy must a quasar wind be to create emission line blueshifts?
}


\author[J. H Matthews]{James~H.~Matthews$^{\orcidlink{0000-0002-3493-7737}}$\thanks{james.matthews@physics.ox.ac.uk}
\\Department of Physics, Astrophysics, University of Oxford, Denys Wilkinson Building, Keble Road, Oxford OX1 3RH, UK\\
}

\date{Accepted 2026 July 24. Received 2026 June 26; in original form 2026 May 21}

\pubyear{2026}

\label{firstpage}
\pagerange{\pageref{firstpage}--\pageref{lastpage}}
\maketitle

\begin{abstract}
Blue asymmetries (``Blueshifts'') in the C~\textsc{iv}~1550\AA\ emission line are common in luminous quasars. If they are formed in winds, how much energy, momentum and mass do those winds transport? We address this question 
by considering how much mass must be supplied through the line-forming region to maintain a given density and ionization state. Using a combination of 1D analytic and 2D numerical models, we find that for blueshifted C~\textsc{iv} lines to form in a wind, the wind must have mass outflow rates of $\sim 50 f_V$ times the accretion rate, where $f_V \leq 1$ is the volume filling factor accounting for clumping. Our results therefore disfavour line formation in a smooth disc wind and point towards one of two scenarios: either the wind is clumpy, with required clumping factors suggestively close to those in hot star winds; alternatively, if the mass is instead swept up from the ambient medium, the wind need not be clumpy and MHD and radiative winds can provide the original source of momentum and energy. 
The power of the outflow depends on the square of the terminal velocity of the flow, $v_\infty$. If the wind is also the BAL outflow, with $v_\infty \sim10,000~{\rm km~s}^{-1}$, the wind power is significant and important for feedback. 
There are various caveats which moderate our conclusions, motivating i) a better theoretical understanding of wind driving and clump formation physics and ii) improved observational constraints on the physical conditions where the lines are formed. 
\end{abstract}

\defcitealias{richards_unification_2011}{R11}
\defcitealias{rankine_bal_2020}{R20}
\defcitealias{matthews_stratified_2020}{M20}
\defcitealias{matthews2023}{M23}
\newcommand{\civ}{\ion{C}{iv}}
\newcommand{\civline}{\ion{C}{iv}~1550\AA}
\newcommand{\sirocco}{\textsc{Sirocco}}

\section{Introduction}
Accreting systems ubiquitously drive outflows, whether they are collimated jets or mass-loaded, wide-angle winds. Often, both jets and winds have a close relationship with the accretion process  \citep[e.g.][]{fender_towards_2004,ponti_ubiquitous_2012,rankine_bal_2020,temple2023,Parra2024,jackson2026}, meaning they are important to understand if we are to build a holistic picture of accretion onto compact objects. In (at least) quasars and other active galactic nuclei (AGN), outflows are important feedback agents, allowing momentum and energy to be transported far from where the accretion energy is originally liberated.
While there are ``many routes to AGN feedback'' \citep{morganti_many_2017}, mass-loaded winds (accretion disc winds and/or larger-scale, cooler outflows) are, one way or another, fundamental to our understanding of how galaxies and black holes co-evolve \citep[e.g.][]{silk_quasars_1998,Fabian1999,king_black_2003,di_matteo_energy_2005,hopkins_quasar_2010,costa_feedback_2014,morganti_many_2017,harrison_agn_2018,laha_ionized_2021}. Constraining the amount of mass, momentum and energy the winds transport is correspondingly important. 

The smoking-gun signature of outflowing material is blue-shifted, broad absorption along the line-of-sight to a continuum source. This observational kinematic imprint provides unambiguous evidence that material between us and the source is moving towards us. Spectacular examples of this phenomenon can be seen in the ultraviolet and/or optical spectra of accreting white dwarfs \citep{heap_iue_1978,greenstein_rw_1982,cordova_high-velocity_1982,kafka_detecting_2004}, X-ray binaries \citep{Ioannou2003,munoz-darias_regulation_2016,fijma2023,castro2022,Castro2026}, FU Orionis systems \citep{bastian1985,hartmann1996,milliner_disc_2019} and BAL quasars \citep{hewett2003,weymann_comparisons_1991,rankine_bal_2020}. In addition, wind signatures can be seen in the X-ray band in XRBs \citep[e.g.][]{ponti_ubiquitous_2012,Parra2024} and AGN \citep[e.g.][]{pounds_high-velocity_2003,gofford_suzaku_2013,tombesi_unification_2013,laha_warm_2014}, where, in the latter case, they comprise both warm absorbers and so-called ultra-fast outflows (UFOs). In quasars\footnote{We take `quasar' to mean an AGN luminous in the rest-frame optical and UV, with $L_{\rm bol} \gtrsim 10^{45}~{\rm erg~s}^{-1}$.}, disc winds form important constituent parts in many unified models for explaining their observational phenomenology \citep[e.g.][]{murray_accretion_1995,elvis_structure_2000,giustini}.

Blueshifts or blue asymmetries in emission lines are also relatively common, especially so in quasars. In particular, the \civline\ has been known to be blueshifted\footnote{What is referred to as a blueshift throughout this paper (which follows most of the related literature) does not correspond to a simple translation of the entire line-profile, but an asymmetry with more flux in the blue wing than the red.} in some quasars since the early 1980s \citep{gaskell_redshift_1982,wilkes_studies_1984}. While they are a less clear cut indicator of outflows or winds, it has long been suspected that they are associated with outflowing material. 

\cite{richards_unification_2011} established the fundamental nature of the \civline\ equivalent width versus blueshift parameter space (``\civ\ emission space''), showing that strongly blueshifted \civ\ emission lines are associated with low equivalent widths and weak \ion{He}{ii}~1640\AA\ emission. Subsequently, \cite{rankine_bal_2020} used a mean-field independent component analysis to characterise the \civ\ emission space and show that BAL and non-BAL quasars were drawn from the same underlying population and that the quasars with the strongest blueshift are also highest Eddington fraction, with, in general, the broadest and fastest BAL troughs. \cite{temple2023} expanded on this further by examining emission line demographics as a function of black hole mass and Eddington fraction, showing good qualitative agreement with accretion and outflow models driven by the underlying spectral energy distribution (SED). A number of other authors have highlighted the fundamental nature of \civ\ emission space by exploring relationships between \civ\ and \ion{He}{ii} emission line properties and other diagnostics from radio \citep{richards_unification_2011,rankine_placing_2021,richards_probing_2021,petley_connecting_2022} and X-ray/extreme-UV wavelengths \citep{gallagher2005,richards_unification_2011,Hiremath2025,Shlentsova2026,Rankine2026}. All in all, it is clear that blueshifted \civ\ emission lines are common, and fundamental in the sense that they are closely related to many of the key quasar observables. Most models interpret the blueshifts as originating in an outflow or disc wind (\citealt{wilkes_studies_1984,Baldwin1996,richards_broad_2002,Leighly2004a,Leighly2004b,yong_kinematics_2017,matthews2023}; although see also \citealt{gaskell_case_2016}), possibly within a line-driven wind framework \citep[e.g.][]{richards_unification_2011,giustini,temple2023}. However, the detailed physics of these putative winds is not well-known and their connection to other outflow phenomena (BALs, UFOS, etc.) remains unclear. 

In this paper, we examine the conditions needed to form blueshifted \civ\ emission lines in quasar spectra and what this implies about disc wind properties. The crux of the argument is fairly simple: to produce a blueshifted emission line, one must be able to keep material moving at a certain velocity towards the observer, and the material must not be over-ionized. A certain density must therefore be maintained, which immediately implies a mass flow rate and corresponding outflow power. With a rough yet reasonable set of assumptions, it turns out that one can derive rather simple predictions for the properties of the putative outflow -- which is exactly the approach here. Similar principles have been in the literature in various contexts, ranging from back-of-the-envelope analytic estimates to detailed modelling. To give four specific examples: i) \cite{ponti_ubiquitous_2012} estimate the mass-loss rates of hot X-ray binary winds based on the ionizaton state of the observed Fe lines; ii) \cite{Choi2020,Choi2022,Choi2022b} use the spectral synthesis code {\sc SimBAL} to constrain the physical properties and associated powers of BAL outflows; iii) Arav and collaborators \citep[e.g.][]{Borguet2013,Chamberlain2015,arav2018,arav2020,Dehghanian2025,Sharma2025} use density sensitive line diagonistics combined with ionization arguments to constrain the physical size scales and kinetic powers of BAL quasars; iv) 
\cite{drew1982,drew1984} use photoionization calculations within a spherically symmetric wind to model BALs, suggesting quasar mass-loss rates on the order of $40-100~M_\odot~{\rm yr}^{-1}$ for their homogenous wind model. 

Here, we apply our specific ionization-based argument using blueshifted emission lines in quasar spectra as our observational tracer. In addition, we extend the method by parametrising the impact of SED shape, clumping, absorption and geometry, then, importantly, {\em calibrating} our estimates using Monte Carlo radiative transfer and photoionization calculations with the \sirocco\ code. We start with a simple quasi-spherical geometry (section~\ref{sec:spherical}), before interrogating the assumptions made using results from more detailed \sirocco\ simulations, which factor in the non-spherical and stratified nature of the quasar disc wind (section~\ref{sec:sirocco}). In section~\ref{sec:conservation}, we estimate the maximum wind efficiency from conservation of mass, momentum and energy, before discussing wind driving mechanisms (section~\ref{sec:driving}). Finally, we discuss the results further and conclude in section~\ref{sec:discuss}.

\section{Line formation in a quasi-spherical wind}
\label{sec:spherical}
We begin by following various previous authors \citep[e.g.][]{reynolds_constraints_2012,ponti_ubiquitous_2012,costa_feedback_2014} in assuming a quasi-spherical wind model in steady-state with constant wind velocity $v_w$. The wind has a mass-loss rate $\dot{M}_{w}$, a momentum transfer rate $\dot{M}_w v_w$ and a kinetic power $\dot{E}_w = \frac{1}{2} \dot{M}_w v_w^2$. We assume that the wind covers $\Omega$ steradians in solid angle such that the covering factor is $\Omega/4\pi$. The Hydrogen number density at any radial coordinate $R$ in the wind follows from mass continuity and is given by 
\begin{equation}
    n_H = \frac{\dot{M}_w} {\Omega R^2 \mu m_p v_w},
    \label{eq:nh1}
\end{equation}
where $\mu$ is the mean atomic weight. This equation is valid for a smooth flow, but if the wind instead is clumpy with some volume filling factor $f_V$ then the density is enhanced by a factor $f_V^{-1}$, giving
\begin{equation}
    n_H = \frac{\dot{M}_w} {f_V \Omega R^2 \mu m_p v_w}.
    \label{eq:nh2}
\end{equation}

\subsection{Wind efficiency parameter}
We define a wind efficiency parameter or mass-loading parameter
\begin{equation}
\epsilon_w = \frac{\dot{M}_w}{\dot{M}},
\label{eq:epsilon_w1}
\end{equation}
where $\dot{M} \equiv L_{\rm bol} / (\eta c^2)$ is the accretion rate defined from the bolometric accretion luminosity, with a radiative efficiency $\eta$. In a steady-state disc without mass-loss, $\dot{M}$ the same as the mass accretion rate at a given point in the disc, but in the case of mass-loss or a disc out of steady state this can be different from the accretion rate at a given cylindrical radius, $\dot{M}_{\rm acc} (r_{\rm cyl})$. The wind efficiency
can be related to the wind properties as 
\begin{equation}
\epsilon_w  = \frac{f_V \Omega R_0^2 \mu m_p v_w n_H}{\dot{M}}.
\label{eq:epsilon_w2}
\end{equation}
Given the definition of $\dot{M}$, we can then write the wind efficiency as 
\begin{equation}
\epsilon_w  = \frac{f_V \Omega R_0^2 \mu m_p v_w n_H}{L_{\rm bol}} \eta c^2.
\label{eq:epsilon_w3}
\end{equation}

\begin{figure}
    \centering
    \includegraphics[width=\linewidth]{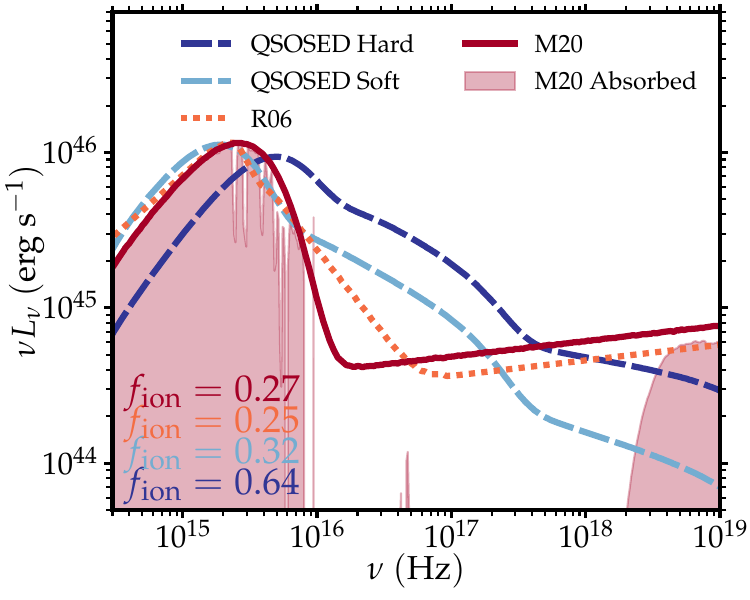}
    \caption{Spectral energy distributions (SEDs) used in this work. R06 (yellow dashed line) is the mean quasar SED from \protect\cite{richards_sloan_2006} and M20 (red solid line) is the disc$+$power-law SED from \protect\citetalias{matthews_stratified_2020}. The red-filled area shows a highly absorbed version of this SED, also used in the {\protect\sc Cloudy} simulations. The two blue lines show SEDs from {\sc Qsosed} \protect\citep{kubota}, representing extremes of `hard' and `soft' from the \protect\cite{temple2023} simulation grid. In each case, the value of $f_{\rm ion}$, the fraction of H-ionizing luminosity (equation~\ref{eq:fion}), is labelled in the same colour.}
    \label{fig:seds}
\end{figure}

\subsection{Estimate from ionization parameter}
\label{sec:ion_est}
Let us now assume that the CIV line forms at a given ionization parameter $\xi$ which is given by (in the optically thin limit)
\begin{equation}
\xi_{\rm thin} = \frac{L_{\rm ion}}{n_H R^2}
\label{eq:xi_thin}
\end{equation}
where $L_{\rm ion}$ is the ionizing luminosity, integrated over the conventional photon energy range of $13.6-1000~\mathrm{eV}$ \citep{tarter_interaction_1969}. We will parameterise absorption of ionizing photons (by an "attenuation" factor $f_{\rm att}$) and relate the ionizing luminosity to the Bolometric luminosity as $L_{\rm ion} = f_{\rm ion} L_{\rm bol}$, such that our ionization parameter becomes 
\begin{equation}
\xi = f_{\rm att} f_{\rm ion} \frac{L_{\rm bol}}{n_H R^2}.
\label{eq:xi}
\end{equation}
$f_{\rm att}$ is thus also equal to $\xi/\xi_{\rm thin}$ and can be written formally in terms of the local mean intensity, $J_\nu$, as 
\begin{equation}
f_{\rm att} = \frac{R^2}{L_{\rm ion}} 
\int_{13.6\mathrm{eV}/h}^{1000\mathrm{eV}/h} 
4 \pi J_\nu d\nu.
\label{eq:fatt}
\end{equation}
The definition of $f_{\rm ion}$ is 
\begin{equation}
f_{\rm ion} = \frac{1}{L_{\rm bol}} 
\int_{13.6\mathrm{eV}/h}^{1000\mathrm{eV}/h} 
L_\nu d\nu,
\label{eq:fion}
\end{equation}
meaning it is only dependent on the shape of the quasar SED. $f_{\rm ion}$ ranges from $0.25-0.64$ for our adopted quasar SEDs as shown in Fig.~\ref{fig:seds}. 
The ionization parameter $\xi$ can be plugged into our expression for the wind efficiency (equation~\ref{eq:epsilon_w3}) to give 
\begin{equation}
\epsilon_w  =\frac{\Omega \mu m_p v_w}{\xi} \eta c^2 f_{\rm att} f_{\rm ion} f_V, 
\label{eq:eps_w_novals}
\end{equation}
which explicitly relates the wind efficiency / mass loss rate to properties we can feasibly infer from quasar spectra (namely $\xi$, $v_w$, $\lambda_{\rm Edd}$) -- albeit with a lot of uncertainty in the other `fudge' parameters. Equation~\ref{eq:eps_w_novals} is similar to equation 1 of \cite{ponti_ubiquitous_2012}, except that i) we have parametrised additional physics, and ii) using a wind efficiency allows us to eliminate the luminosity dependence. 

\begin{figure*}
    \centering
    \includegraphics[width=\linewidth]{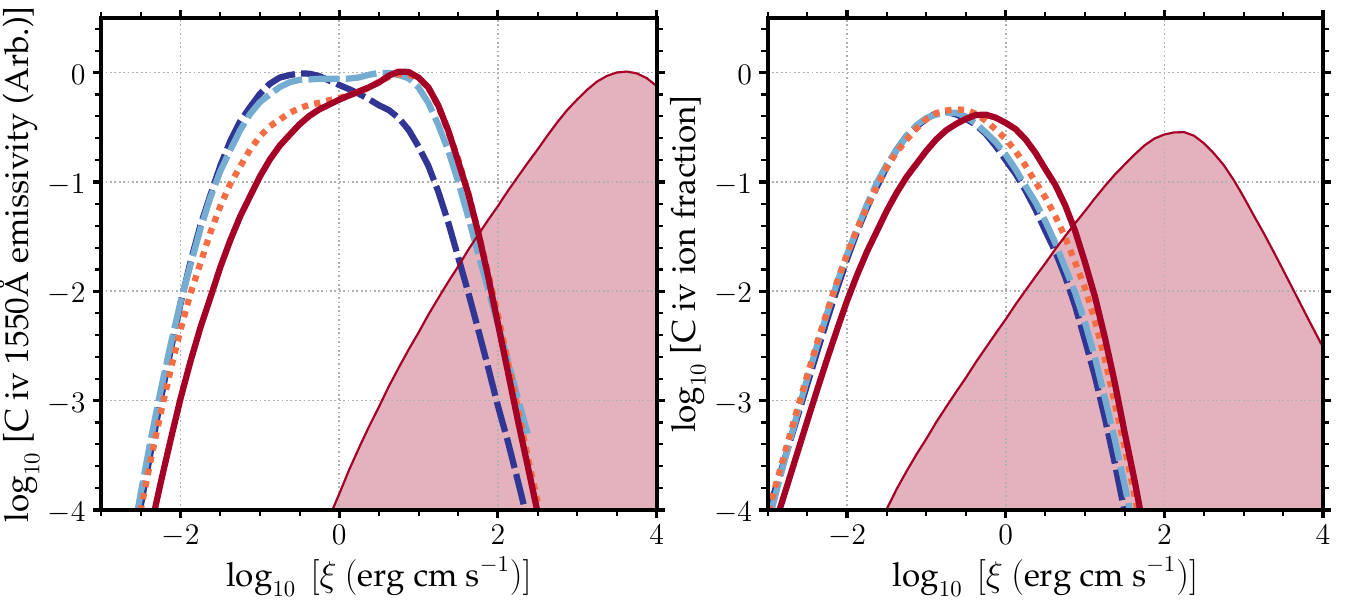}
    \caption{Ionization parameter of \civline\ line formation in our {\sc Cloudy} models, for each of the SEDs shown in Fig.~\ref{fig:seds}. In the left-hand panel we show the \civline\ line emissivity, normalised to the maximum value, wherea in the right hand panel we instead show the \civ\ ion fraction. The unabsorbed SEDs show relatively similar results albeit with a range of peak values, whereas the extremely absorbed SED (filled-red area) allows \civ\ to form at much higher $\xi$. 
    }
    \label{fig:cld}
\end{figure*}

To obtain our mass-loading efficiency estimate, we need an estimate of $\xi$. We start by using \textsc{Cloudy} \citep{ferland17} to gain a handle on the characteristic ionization parameter of \civline\ formation. Specifically, we consider a single cloud of column density $N_H = 10^{19}~{\rm cm}^{-2}$ and density $n_H = 10^{11}~{\rm cm}^{-3}$ illuminated by the same four SEDs shown in Fig.~\ref{fig:seds}. In Fig.~\ref{fig:cld}, we show the resulting ion fraction and \civline\ line emissivity in the cloud as a function of $\log \xi$. The peak of the line emissivity depends a little on the specific SED and the dependence on $\log \xi$ can be rather flat near the peak. 
The simulations suggest ionization parameters of $-1 \lesssim \log \xi \lesssim  1$  in the region of the BLR where \civline\ is emitting most efficiently. 

Absorption, however, can dramatically affect the transmitted SED and the ionization parameter at which a given ion is abundant. To assess the potential impact of absorption, we consider the heavily absorbed M20 SED from Fig.~\ref{fig:seds}, which has almost no flux between $10^{16}~{\rm Hz}$ and $10^{18}~{\rm Hz}$ (41.4 and 4140 eV), a region which contains the \ion{C}{iii}$\to$\ion{C}{iv} and \ion{C}{iv}$\to$\ion{C}{v} ionization edges. The red shaded region in Fig~\ref{fig:cld} then shows the corresponding \civ\ ion abundance curve with this SED, whose peak is shifted to higher $\xi$ by around $2.5$~dex. 

This large $\xi$--offset can be understood by considering the photoionization physics. The photoionization rate from a given ion, $i$, with a threshold energy $h \nu_i$ is given by
\begin{equation}
    \gamma_i = \int_{\nu_i}^{\infty}\frac{4 \pi J_\nu \sigma_{i}(\nu)}{h\nu}~d \nu \, ,
\end{equation}
where $\sigma_{i}(\nu)$ is an appropriately weighted total cross-section, bearing in mind that the true photoionization rate would be obtained from sums over each level or configuration of the ion. The equation makes clear that the photoionization rate depends on the ion in question through its threshold energy and the shape of its photoionization cross-section. Cross-sections typically drop quite steeply with photon energy, with hydrogenic cross-sections going as $\nu^{-3}$. As a result, the true {\em ionizing efficacy} of a given SED is strongly frequency-weighted by the cross-section, and is most sensitive to the radiation close to the relevant ionization edge. 
In our absorbed SED, there is a lot of flux at very energies and just above the Lyman edge, but almost no flux around the relevant Carbon edges. As a result, the \ion{C}{iv} ion forms at much higher $\xi$ than when using the unabsorbed SED, with the offset approximately given by the ratio of SED ionizing efficacies. 

Our absorbed SED is fairly extreme, and so the appropriate estimate for $\xi$ is likely to be bounded by the maxima of the curves in Fig.~\ref{fig:cld}. Indeed, \cite{Leighly2004b} investigate a more modest shielding effect and find correspondingly less drastic effects. Furthermore, although these \textsc{cloudy} models are instructive, as they are 1D, they do not capture the full range of radiative transfer and self-shielding effects or density gradients within a quasar disc wind. In our MCRT and photoionization simulations of disc winds, which do model these effects in full, the \civline\ can be formed over a range of physical conditions as discussed in section~\ref{sec:results_xi}. There, we do indeed find an intermediate regime, with \civ\ line emission produced from a range in $\xi$ of around one decade, such that $\log \xi \sim 1.5-2.5$, with a wide range of \civ\ line emissivities at a given $\xi$. Taking $\log \xi \sim 2$ as a typical value for the line formation region, we find 
\begin{equation}
\epsilon_w  \sim 
50~
f_V 
\left(\frac{\Omega}{\pi}\right)
\left(\frac{v_w}{3000~{\rm km~s}^{-1}}\right)
\left(\frac{\xi}{100}\right)^{-1}
\left(\frac{f_{\rm ion}}{0.25}\right)
\left(\frac{f_{\rm att}}{0.1}\right).
\end{equation}
It is worth noting at this point that the above equation can feasibly be evaluated for actual quasar spectra, assuming values for $\Omega$, $\xi$ and $f_{\rm att}$. In principle, one can make estimates of the ratio $\epsilon_w/f_V$ for quasar winds as a function of, e.g., $M_{\rm BH}$ and $L/L_{\rm Edd}$, which seems useful. We note that this estimate is broadly consistent with that obtained from modelling of \civ\ BALs by \cite{drew1982}.

\subsection{Estimate from emission line ratios}
\cite{temple_high-ionization_2021} used photoionization calculations with Cloudy, applied to quasar spectra, to show that the variation in line ratios in quasars could be explained by changes in density and ionizing flux, rather than any change in metallicity. In particular, \cite{temple_high-ionization_2021} find that in `wind-dominated' quasar spectra -- those with relatively weak emission lines with strong blueshifted components -- the \ion{N}{v}/Ly-$\alpha$, \ion{N}{v}/\ion{C}{iv} and (\ion{Si}{iv}+\ion{O}{iv})/\ion{C}{iv}  ratios are high compared to those in `core-dominated' systems. The results can be explained if the `outflowing' (i.e. blueshifted) broad emission lines are preferentially associated with dense gas with $n_H \sim 10^{12-14}~{\rm cm}^{-3}$ and high fluxes of ionizing photons, $\phi_H \sim 10^{23}~{\rm cm}^{-2}~{\rm s}^{-1}$. The photon flux can be related to $L_{\rm bol}$ for a given SED shape, by 
\begin{equation}
    \phi_H = \frac{L_{\rm bol}}{4\pi R^2} \left(\frac{Q_H}{L_{\rm bol}}\right) g_{\rm att} \, .
\label{eq:phi_H}
\end{equation}
Here the first term describes the geometric dilution of the radiation field, the second $Q_H/L_{\rm bol}$ term has units of erg$^{-1}$ and represents the number of ionizing photons produced per unit energy released. The parameter $g_{\rm att}$ is the fraction of ionizing photons that are absorbed/attenuated by opacity sources between the ionizing source and the line formation region. 
The second term is given by 
\begin{equation}
\frac{Q_H}{L_{\rm bol}} = \frac{1}{L_{\rm bol}} \int_{13.6{\rm eV}/h}^{\infty} \frac{L_\nu}{h\nu} d\nu \, .
\end{equation}
Our unabsorbed SEDs have $Q_H/L_{\rm bol}$ in the range $(1-3)\times10^{10}~{\rm erg}^{-1}$, so we adopt $10^{10}~{\rm erg}^{-1}$. Similar $Q_H/L_{\rm bol}$ values are found by \cite{wilkins2025} for their {\sc Qsosed} models. If we now assume $g_{\rm att}=0.1$, we can rearrange equation~\ref{eq:phi_H} to obtain a radius of line formation
\begin{equation}
R \sim 3\times10^{15}~{\rm cm} 
\left(\frac{L_{\rm bol}}{10^{46}}\right)^{1/2}
\left(\frac{g_{\rm att}}{0.1}\right)^{1/2}
\left(\frac{\phi_H}{10^{23}~{\rm cm}^{-2}~{\rm s}^{-1}}\right)^{-1/2} \, .
\end{equation}
Putting this into equation~\ref{eq:epsilon_w3} we obtain 
\begin{equation}
\epsilon_w \sim 157~f_V~
\left(\frac{\Omega}{\pi}\right)
\left(\frac{\phi_H}{10^{23}~{\rm cm}^{-2}~{\rm s}^{-1}}\right)^{-1}
\left(\frac{n_H}{10^{12}~{\rm cm}^{-3}}\right)
\left(\frac{g_{\rm att}}{0.1}\right),
\end{equation}
which is independent of $L_{\rm bol}$, remembering that $\phi_H$ is the local ionizing photon flux required to produce the observed line ratios. In this work, we hereafter focus on the ionization parameter estimate from equation~\ref{eq:epsilon_w1}. However, we note i) the general potential of this alternative method for constraining mass-loading efficiencies, and ii) that the estimate agrees with that from equation~\ref{eq:epsilon_w1} within a factor of $\approx 3$. 

\subsection{Interrogating the assumptions}

In the preceding analysis, there are three key assumptions to be examined, all of which are related:
\begin{enumerate}
\item {\bf In equation~\ref{eq:epsilon_w1}, 
we have assumed that the \civline\ line forms at a single characteristic ionization parameter, $\xi$}. This approach is simplistic, for three reasons. First, because the `ionization state' is not a single number, but is  described by a vector of ion populations, $\vec{{\cal N}}_{\rm ion}$. Technically, for any given $\xi$ and $n_H$ there are an infinite number of possible SEDs, each 
with a corresponding $\vec{{\cal N}}_{\rm ion}$. Furthermore, the SED varies throughout the wind (see point ii.). Second, because any given line can form over a range of $\xi$; this behaviour is illustrated by the non-neglible width of the line emissivity curve in Fig.~\ref{fig:cld}, as well as the corresponding width of contours in plots of line emissivities over $\phi_H--n_H$ space \citep[e.g.][]{korista_atlas_1997,matthews_stratified_2020,temple_high-ionization_2021}, where a constant ionization parameter (in this case $U$) makes a diagonal line. Third, because $\xi$ is the {\em Hydrogen} ionization parameter, with a threshold energy of $13.6~{\rm eV}$, whereas the relevant edges for creation (\ion{C}{iii}$\to$\ion{C}{iv}) and destruction (\ion{C}{iv}$\to$\ion{C}{v}) for the \ion{C}{iv} ion are  47.88~eV and 64.49~eV, respectively. 

\item {\bf We have crudely approximated the multi-dimensional radiative transfer and self-shielding effects through the single parameter $f_{\rm att}$.} In reality, the ionization conditions are determined by the local mean intensity, which depends on the frequency-dependent transport of radiation through the stratified wind structure, which is in turn dependent on opacities and emissivities that in turn depend on the local ionization and excitation state. 

\item {\bf We have assumed a quasi-spherical coasting wind, such that $n_H \propto R^{-2}$}. The more general form of mass continuity follows from considering the mass contained in a volume element $dV = v_w dA_\perp dt$, where $dA_\perp$ is the cross-section area perpendicular to the streamline. The number density follows as $n_h = (\mu m_p v_w)^{-1} dM/(dA_\perp dt)$. In a true quasi-spherical geometry, i.e. one where streamlines are radial and converge at the origin, the cross-sectional area can then be related to solid angle as $dA_\perp = R^2~d\Omega$; for uniform mass-loss per unit solid angle one then recovers equation~\ref{eq:nh1}.  In more realistic disc wind geometries, the wind forms a hollow bicone and is launched from a disc rather than a single point. The quasi-spherical limit will be approached asymptotically at large distances, but close to the disc the geometry differs substantially, and the $dA_\perp \propto R^{2}$ dependence that drives the $n_H \propto R^{-2}$ relationship breaks down. More generally, any quasar BLR or outflow is at minimum 2D in nature, so is subject to both the multi-dimensional radiative transfer and self-shielding effects mentioned above as well as these non-radial density gradients.
\end{enumerate} 
On the one hand, these assumptions are within the spirit of approximation of the calculation and allow one to obtain a simple yet instructive relationship between $\xi$ and the wind parameters. On the other, all of them are over-simplifications. We now improve on them by carrying out 2.5D Monte Carlo radiative transfer simulations. 

\section{2D Disc Wind Models with \sirocco}
\label{sec:sirocco}

We now dispense with the above approximations and consider a more realistic 2D geometry, treated with accurate Monte Carlo radiative transfer simulations. The aim is not to replace the previous analytic estimate, rather to establish whether a more complete approach relaxes the requirement for highly mass-loaded and/or clumpy winds (that is, large values of $\epsilon_w / f_V$). 

We consider a 2D disc wind in which the wind is azimuthally symmetric around the polar axis and has reflection symmetry about the midplane. The wind forms a hollow bicone rising from the disc, a geometry typical of disc wind models \citep[e.g][]{emmering_magnetic_1992,murray_accretion_1995,elvis_structure_2000,sim_multidimensional_2008,sim_multidimensional_2010,hagino_origin_2015}. Specifically, we adopt the geometry of \cite{shlosman_winds_1993}, originally applied to accreting white dwarfs, which has since been used a number of times to model quasar disc winds \citep{higginbottom_simple_2013,yong_black_2016,yong_kinematics_2017,matthews_testing_2016,matthews_stratified_2020,matthews2023}. This provides a flexible prescription with adjustable wind geometry, launching radius and kinematics. 

We use the Monte Carlo radiative transfer and ionization code \sirocco\ \citep{long_modeling_2002,sirocco} to calculate the physical properties of the flow for the \cite{shlosman_winds_1993} parametrised wind geometry. We first run 20 `ionization cycles' to iteratively calculate the ionization and temperature structure of the wind -- this is done by flying $N_\gamma = 1.5\times10^7$ photon packets, or $r$-packets, through a discretised wind grid and recording their ionizing and heating impact through Monte Carlo estimators. The $r$-packets are initialised so that their frequency distribution and energy weights reproduce the desired input quasar SED. Once the wind plasma state is approximately converged, we compute the emergent rest-frame UV spectrum (around the \civline\ line) at a range of viewing angles. This exercise then allows us to examine whether a given wind geometry can give rise to a significant \civline\ emission feature -- this is a minimal requirement for a \civ\ blueshift under a disc wind formation hypothesis, and allows us to assess the impact of our assumptions in the simplified quasi-spherical model. 

\subsection{Simulation Setup}
We run a simulation grid of 200 \sirocco\ simulations. We consider a $10^9~M_\odot$ black hole and the same illuminating SED as \citet[][hereafter \citetalias{matthews_stratified_2020,matthews2023}]{matthews_stratified_2020,matthews2023}, shown in Fig.~1, which is a combination of a multi-temperature disc blackbody and an X-ray power law with $F_\nu \propto \nu^{-0.9}$. The $r$-packets are launched isotropically from the origin, which is a conservative assumption since the strength of the blueshifted \civline\ decreases if disc anisotropy is introduced \citepalias{matthews_stratified_2020,matthews2023}. We normalise our SED to produce a bolometric luminosity of $L_{\rm bol} = 2.94 \times 10^{46}~{\rm erg~s}^{-1}$, corresponding to a mass accretion rate of $\dot{M}\approx 5~M_\odot~{\rm yr}^{-1}$ for a radiative efficiency of $\eta = 1/12$, and an Eddington fraction of $L_{\rm bol} / L_{\rm Edd} \approx 0.2$. The 2-10 keV X-ray luminosity is $10^{45}~{\rm erg~s}^{-1}$, corresponding to an optical to X-ray spectral index of 
$\alpha_{\rm ox} = 0.3838 \log_{10} (L_{2~{\rm keV}} / L_{2500}) = -1.43$. 
Overall, the energetics, scale lengths and basic observable properties are reasonable for a quasar exhibiting blueshifted quasar emission lines \citep{temple2023}. The SED shape is more uncertain, but the range of $f_{\rm ion}$ values for the SEDs in Fig.~\ref{fig:seds} is not particularly large, and, as shown by \citetalias{matthews_stratified_2020,matthews2023}, alternative reasonable SED choices do not dramatically alter the emergent spectrum from comparable MCRT models. 

With the basic quasar properties specified, we then vary the following parameters by Monte Carlo sampling each using the designated strategy:
\begin{itemize}
    \item $\dot{M}_w \in (0.01,100)$: the mass loss rate of the wind in $M_\odot~{\rm yr}^{-1}$ (log-uniform sampling)
    \item $f_V \in (0.001,1)$: the volume filling factor of the wind (log-uniform sampling)
    \item $\theta_{\rm min} \in (20^\circ, 70^\circ)$: the inner opening angle of the wind bicone (uniform sampling)
    \item $R_v \in (10^{17}~{\rm cm}, 10^{19}~{\rm cm})$: the wind acceleration length (log-uniform sampling)
    \item $r_{\rm min} \in (60~r_g, 1500~r_g)$: the inner launch radius of the wind, where the outer launch radius is set to $r_{\rm max} = 2 r_{\rm min}$ (uniform sampling).
    \item $\alpha \in (0.5,2)$: the acceleration exponent of the wind (uniform sampling)
\end{itemize}
For each run, the value of each parameter is chosen randomly from the quoted range with the designated sampling strategy. We seek to ascertain whether our adopted critical $\epsilon_w / f_V$ value is reasonable; this approach gives a set of simulations with a range of values of $\epsilon_w$ and $f_V$ while treating the remaining parameters ($\theta_{\rm min}$, $R_v$, $\alpha, r_{\rm min})$ analogously to nuisance parameters. 

\subsection{Results}
\label{sec:results}

\begin{figure}
    \centering
    \includegraphics[width=\linewidth]{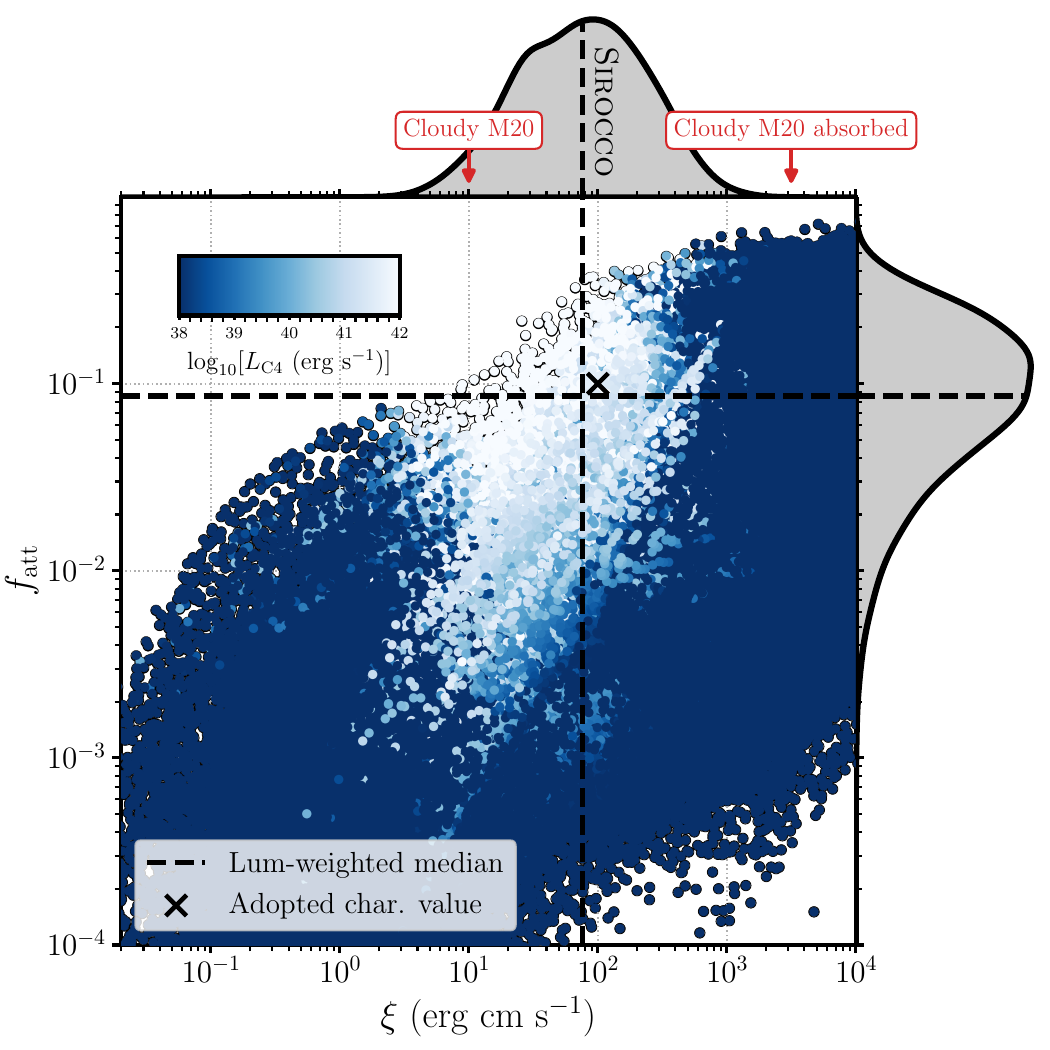}
    \caption{The ionization--attenuation ($[\xi,f_{\rm att}]$) parameter space, showing under what conditions the \civline\ forms efficiently. In the main panel, we show the shielding/attenuation parameter $f_{\rm att}$ (equation~\ref{eq:fatt}) plotted as a function of $\xi$ from all 200 \sirocco\ simulations, colour-coded by the \civline\ line luminosity, $L_{\rm C4}$. The bright region shows where \civline\ forms efficiently. The curves on the axes show the 1D weighted distributions of each parameter, obtained from a kernel density estimate with $L_{\rm C4}$ as the weight. The dashed lines show the weighted medians of $\log \xi = 1.88$  and $f_{\rm att} = 0.086$, which are close to the characteristic values adopted in section~\ref{sec:ion_est}. The red arrows show the peak of the $L_{\rm C4}$ distributions from our {\sc Cloudy} simulations from Fig.~\ref{fig:cld}, which bound the \sirocco\ estimates.}
    \label{fig:fatt}
\end{figure}

\begin{figure*}
    \centering
    \includegraphics[width=0.7\linewidth]{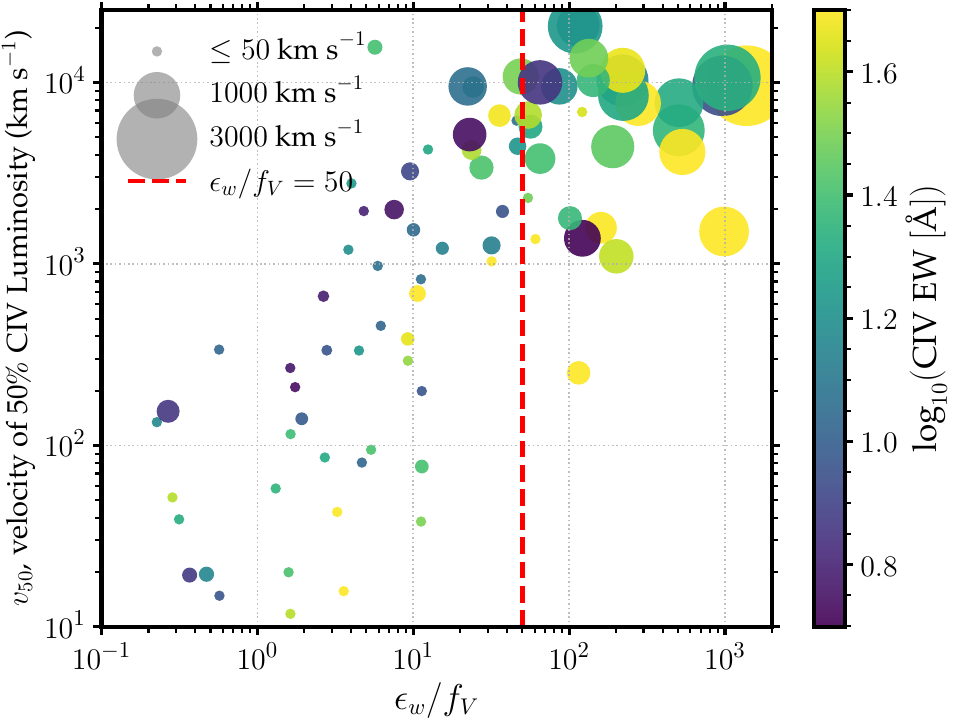}
    \caption
    {
    High values of $\epsilon_w f_V^{-1}$ are required to produce large blueshifts and velocities. The plot shows $v_{50}$, the poloidal velocity at which the CDF of the \civ\ line luminosity is equal to 0.5, as a function of $\epsilon_w f_V^{-1}$ for our full simulation grid, where $\epsilon_w$ and $f_V$ are the wind efficiency parameter or mass-loading parameter (equation~\ref{eq:epsilon_w1}), and volume filling factor, respectively. $v_{50}$ is a weighted median that can be thought of as the characteristic poloidal velocity of the \civ\ line formation region (see text). The size of the points encodes the blueshift defined through equation~\ref{eq:blueshift}, whereas the colour shows the equivalent width (EW), both at a $10^\circ$ viewing angle. Only the 83/200 runs which produce \civ\ significant emission line luminosities are shown. 
    }
    \label{fig:vel_v_c4lum}
\end{figure*}

\subsubsection{Where does  \civline\ form in $[\xi,f_{\rm att}]$ space?}
\label{sec:results_fatt}
\label{sec:results_xi}

To examine appropriate estimates for the ionization parameter, $\xi$ and shielding parameter, $f_{\rm att}$, we calculate both quantities in our simulations, using a Monte Carlo volume-based estimator for $\xi$; this $\xi$ estimator is a diagnostic, with the ion abundances calculated from the full rate equations following a MC estimator formalism \citep{lucy2003,sirocco}. The parameter $f_{\rm att}$, which parametrises the amount of shielding due to attenuation of ionization photons, is calculated as $\xi / \xi_{\rm thin}$. Fig~\ref{fig:fatt} shows $f_{\rm att}$ as a function of $\xi$ from every cell in all 200 wind models, colour-coded by \civline\ line luminosity, $L_{\rm C4}$. The bright region highlights the area of $[\xi,f_{\rm att}]$ parameter space where the \civ\ line forms effectively, and the histograms show the $L_{\rm C4}$-weighted  distribution of each variable. 

We find that \civline\ forms at a slightly higher $\xi$ in our multidimensional \sirocco\ models compared to the unabsorbed 1D {\sc Cloudy} models discussed in section~\ref{sec:ion_est}. However, the range of $\xi$ from \sirocco\ is also lower than the highly absorbed {\sc Cloudy} model, confirming our suspicion that a realistic shielding scenario will lie somewhere in between the unabsorbed and extreme absorption case (both marked by red arrows). The weighted median of the distribution is $\log \xi = 1.88$, although \civ\ forms over a range of ionization parameters, reflecting the diversity in absorbed SEDs -- both from simulation to simulation and within one simulation due to a stratified ionization structure. The proximity of the weighted median value to our adopted estimate of $\log \xi = 2$ in section~\ref{sec:ion_est} validates it {\sl a posteriori} (and was in fact used to inform said estimate).  

Fig~\ref{fig:fatt}  shows that the \civline\ line luminosity is peaked around $f_{\rm att}\approx 0.1$, with a weighted median of 0.086, although efficient line formation can occur over a range of values down to $f_{\rm att} < 0.01$. Again, the fact that this weighted median validates our choice of estimate ($f_{\rm att}=0.1$) in the analytic model. The $f_{\rm att}$--dependence reveals some interesting behaviour. The lack of line formation at very low $f_{\rm att}$ shows that one cannot simply increase shielding arbitrarily; eventually, there are simply not enough ionizing photons reaching dense wind regions to power luminous line emission. In addition, the plot shows that it is not possible to achieve $\xi \lesssim 100$ in these models without some attenuation or self-shielding in the disc wind, highlighting the critical role of radaitive transfer when making inferences about physical conditions from the ionization state. 

\subsubsection{Which models can produce \civ\ emission line blueshifts?}
\label{sec:results_which}
\cite{coatman2016} and \cite{rankine_bal_2020} define the emission line blueshift as 
\begin{equation}
    {\rm blueshift} = -c\frac{\lambda_{\rm half} - \lambda_0}{\lambda_0} \, ,
\label{eq:blueshift}
\end{equation}
where $\lambda_0=1549.5$\AA\ is the line centre wavelength and $\lambda_{\rm half}$ is the wavelength that bisects the integrated line flux. We do not examine the synthetic spectra in detail here, although in some cases we will use them to verify our methods. For detailed spectral modelling we refer the reader to other studies of emission line blueshifts using \sirocco\ simulations \citepalias{matthews2023} or alternative techniques \citep{chajet_magnetohydrodynamic_2013,chajet_magnetohydrodynamic_2017,yong_kinematics_2017}. The spectra produced are very sensitive to both inclination and the exact kinematic parameters chosen \citepalias{matthews2023}, and a realistic match to quasar emission lines from a self-consistent model is not yet forthcoming. 

Instead, to evaluate the potential for a given model to produce emission-line blueshifts, we adopt an inclination-agnostic method based on the detailed wind cell properties. Our adopted metric corresponds to a weighted median of the poloidal velocity of the wind cells, with \civ\ line luminosity as the weighting. To calculate this we first sort the wind cells by poloidal velocity, $v_l$. Next, from this array, we produce the cumulative distribution function (CDF) of \civline\ line luminosities from all cells, defined as 
\begin{equation}
    {\rm CDF} (v_l) = \frac{\sum_i^{v_i < v_l} L_{\rm C4,i}}{\sum_i L_{\rm C4,i}} \, ,
\end{equation}
where $v_i$ and $L_{\rm C4,i}$ are, respectively, the poloidal velocity and \civline\ luminosity of wind cell $i$. For each wind model we then invert the CDF to find, $v_{50}$, the poloidal velocity at which ${\rm CDF} (v_l) = 0.5$. Cells with $v_l < v_{50}$ therefore contribute to half of the \civline\ line luminosity, meaning $v_{50}$ is a measure of the characteristic poloidal velocity of the \civ\ line formation region. For models with very weak emission lines, $v_{50}$ is not meaningful and can be skewed by MC noise, so we exclude models with $\sum_i L_{\rm C4,i} < 10^{43}~{\rm erg~s}^{-1}$ or \civ\ EW $< 5$\AA\ from our analysis; 83/200 models survive this cut. The excluded models cannot produce observable \civ\ emission line blueshifts and are usually over-ionized with low values of $\epsilon_w f_V^{-1}$. We have verified that $v_{50}$ correlates well with \civ\ blueshift at an inclination, $\theta_i$, of $10^\circ$.

In Fig~\ref{fig:vel_v_c4lum} we plot our blueshift diagnostic, $v_{50}$, as a function of $\epsilon_w f_V^{-1}$ for all considered models. The points are colour-coded by \civ\ EW and the point size represents the blueshift, both evaluated at $\theta_i=10^\circ$. We find a positive correlation between $v_{50}$ and $\epsilon_w f_V^{-1}$ (Spearman rank correlation coefficient of $0.64$), as expected from our 1D analytic estimate. For $v_{\rm 50}$ to exceed $1000~{\rm km~s}^{-1}$, we require $\epsilon_w f_V^{-1} \gtrsim 10$ and all the spectra that show strong blueshifts at $\theta_i=10^\circ$ are concentrated in the upper-right portion of the plot, demonstrating the validity of our $v_{50}$ metric. In fact, all $\theta_i=10^\circ$ spectra with \civ\ blueshifts exceeding $1000~{\rm km~s}^{-1}$ have $\epsilon/f_V \gtrsim 100$. For $\epsilon/f_V < 50$, the mean blueshift is consistent with zero, whereas the maximum is $\approx 680~{\rm km~s}^{-1}$. By contrast, for $\epsilon/f_V > 50$, the mean blueshift is $\approx 800~{\rm km~s}^{-1}$ and the maximum is $\approx 2990~{\rm km~s}^{-1}$. Overall, the numerical results support the central analytic argument from the preceding section, both from a qualitative and quantitative standpoint, so we preserve our estimate of $\epsilon/f_V \sim 50$. In fact, it is possible this estimate is rather conservative for highly asymmetric \civ\ lines. 

We note that the model spectra with the largest blueshifts quite often have relatively large \civline\ EWs. Such a trend might appear to go against the observed one, where large \civ\ blueshifts are only produced when the EW is low \citep{richards_unification_2011,rankine_bal_2020}. As discussed by \citetalias{matthews2023} (their section 5.2), it is probably more reasonable to interpret the observed \civ\ emission space behaviour within a two-component framework, with changes in the relative strengths of `windy' and systemic components cross the parameter space \citep[see also][]{temple_high-ionization_2021}.

In section~\ref{sec:spherical}, we considered five different SEDs for our calculations, but for the \sirocco\ simulations we have considered only one illuminating SED. This is partly for simplicity and to minimise computational cost, and partly based on our previous work: we have shown (see e.g. \citetalias{matthews2023}, their figure 12) that the \civ\ emission line does not dramatically change in luminosity or shape for reasonable choices of illuminating SED. However, our choice is important to interrogate. In reality, we might expect an individual spectrum to be somewhat sensitive to the SED that illuminates the line formation region, and, from an empirical perspective, we know that the magnitude of \civ\ blueshift is closely tied to the illuminating ionizing flux \citep{richards_unification_2011,rankine_bal_2020,temple2023}. Nevertheless, we do not expect our overall science conclusions to depend on this choice, because the general argument outlined in section~\ref{sec:spherical} should apply to a range of SEDs as long as the change in $f_{\rm ion}$ is not too large. To verify this expectation, we ran an additional 40 simulations with the `{\sc Qsosed} soft' SED. We find similar results are produced; all models with high blueshifts are produced in the same region of parameter space as in Fig~\ref{fig:vel_v_c4lum}, and we obtain a similar correlation between $v_{50}$ and $\epsilon_w f_V^{-1}$ (Spearman rank correlation coefficient of $0.54$). Model spectra can show somewhat weaker EWs with the `{\sc Qsosed} soft' SED, but this is largely driven by changes in the continuum under the line, and in any case makes our choice of SED conservative. Thus, while the individual model spectra may change quantitatively with different SED choices, our overall science conclusions are not affected.

\section{Mass, momentum and energy conservation}
\label{sec:conservation}
We now explore the (maximum) wind efficiencies we expect from conserving mass, momentum and energy, considering the disc as the mass source and the radiation field as the momentum and energy source, before examining which mechanisms can drive the wind. 

\begin{figure}
    \centering
    \includegraphics[width=\linewidth]{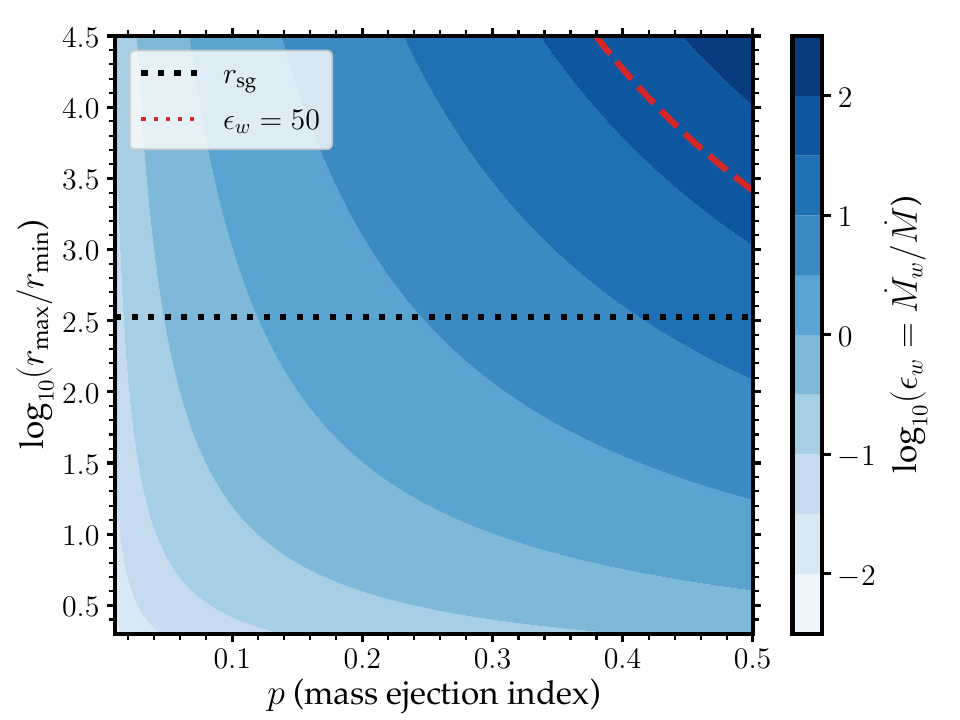}
    \caption{The maximum wind mass-loading for a continually mass-losing disc with $\dot{M}_{\rm acc}\propto r_{\rm cyl}^p$ . The contours show the mass-loading parameter for MHD winds as a function of the ejection index, $p$, and dynamic range in launching radii, $r_{\rm max}/r_{\rm min}$. The black dotted line shows the value of $r_{\rm max}/r_{\rm min}$ for an approximate self-gravitation radius of $R_{\rm sg} = 0.03$~pc. The red dashed line shows $\epsilon_w=50$ from equation~\ref{eq:epsilon_w1}, which requires unrealistically large values of $p$ and $r_{\rm max}/r_{\rm min}$.}
    \label{fig:mhd}
\end{figure}

If the mass comes directly from the disc, a naive expectation for the maximum mass-loss rate is that $\dot{M}_w \sim \dot{M}$, which then implies, very simply, $\epsilon_w \sim 1$. However, if the mass is lost from the outer disc then $\dot{M}_w $ can be higher than $\dot{M}$, since the latter is the accretion rate responsible for the radiated luminosity. Indeed, it is possible to have an accretion rate that is steadily increasing with cylindrical radius $r_{\rm cyl}$ as a result of mass being continually lost. This is the situation within some MHD wind models, as discussed in section~\ref{sec:mhd}, where the mass accretion rate through the disc obeys $\dot{M}_{\rm acc} (r_{\rm cyl}) \propto r_{\rm cyl}^p$, with $p \lesssim 1$ a mass ejection index. In this framework, the total mass loss rate for a wind launched between cylindrical radii $r_{\rm min}$ and $r_{\rm max}$ is \citep{Chakravorty2016}
\begin{equation}
\dot{M}_{\rm wind} = \dot{M}_{\rm acc}(r_{\rm min}) \left[ \left(\frac{r_{\rm max}}{r_{\rm min}}\right)^{p} - 1 \right] \, ,
\label{eq:mdot_wind1}
\end{equation}
which, if we assume that $\dot{M} \approx \dot{M}_{\rm acc}(r_{\rm min})$, gives
\begin{equation}
\epsilon_{w} \approx \left[ \left(\frac{r_{\rm max}}{r_{\rm min}}\right)^{p} - 1 \right]  \, .
\label{eq:mdot_wind2}
\end{equation}
We plot contours of $\epsilon_{w}$ in Fig.~\ref{fig:mhd} as a function of $p$ and the size of the disc, $r_{\rm max}/r_{\rm min}$, with $\epsilon_w = 50$ marked with a red dashed line. We also mark the approximate value of $r_{\rm max}/r_{\rm min}$ implied by a self-gravitation radius of $R_{\rm sg} = 0.03$~pc as estimated from equation~22 of \cite{Collin-Souffrin1990}. Even with the $\dot{M}_{\rm acc} (r_{\rm cyl}) \propto r_{\rm cyl}^p$ assumption, the limit on the size of discs in quasars makes it quite challenging for a disc to lose more mass through a wind than is accreted through the inner disc, but it is possible. However, achieving $\epsilon_w \sim 50$ requires very high values of $p$ approaching unity.

We now consider the case where the rate at which momentum is transported by the flow is equal to the maximum rate of momentum transfer by the radiation field, i.e. $L_{\rm bol} / c$. In this case we can write
\begin{equation}
\dot{M}_w v_w = \frac{\eta \dot{M} c^2}{c}
\end{equation}
which gives a momentum-limited wind efficiency of 
\begin{equation}
\epsilon_w = \frac{\eta c}{v_w} \sim 10~\frac{\eta}{0.1} \left(\frac{\beta_w}{0.01}\right)^{-1} \, ,
\label{eq:zeta_momentum}
\end{equation}
where $\beta_w = v_w/c$ is the wind velocity in units of $c$. The adopted value of $\beta_w = 0.01$ corresponds to a large blueshift of $\approx 3000~{\rm km~s}^{-1}$. 

Finally, we can assume that the kinetic power of the outflow is equal to that of the luminosity of the system, i.e. 
\begin{equation}
\frac{1}{2} \dot{M}_w v_w^2 = \eta \dot{M} c^2
\end{equation}
which gives an energy-limited wind efficiency of 
\begin{equation}
\epsilon_w = \frac{2 \eta  c^2}{v_w^2}\sim 2000~\frac{\eta}{0.1} \left(\frac{\beta_w}{0.01}\right)^{-2} \, ,
\label{eq:energy_limit} 
\end{equation}
where $\beta_w = v_w/c$ is the wind velocity in units of $c$. 

\begin{figure*}
    \centering
    \includegraphics[width=\linewidth]{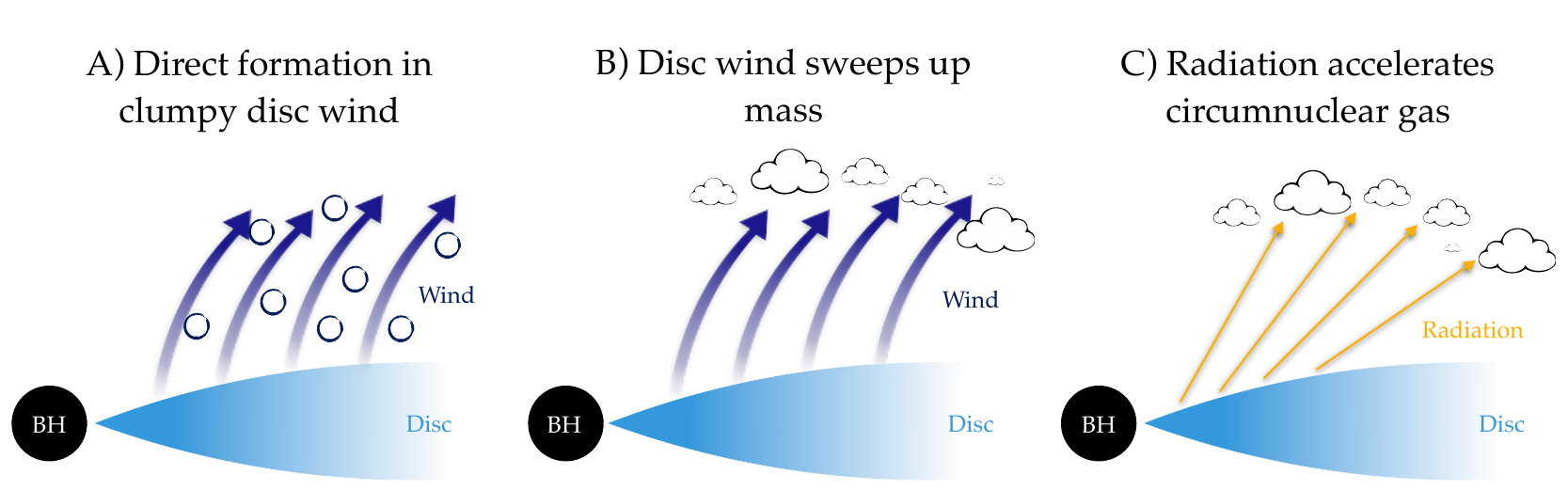}
    \caption{Cartoon illustrating the three possible scenarios for blueshifted broad line formation. In scenarios A and B, the disc wind can be line-driven or magnetically driven. In scenario A, the mass-loss rate is limited at some level by the accretion rate, since the disc directly supplies the mass. In scenarios B and C the mass-loss rate is limited by energetics and other reservoirs are an important source of mass in the outflow.}
    \label{fig:schematic}
\end{figure*}

Consideration of these limits leads us to three possible scenarios that allow the $\epsilon / f_V \sim 50$ condition to be realised, shown in Fig.~\ref{fig:schematic}. the first follows directly from the equation: the disc wind is clumpy with $f_V \sim 10-50$, and $\epsilon_w \sim 1-5$ (scenario A). In this case, the \civline\ line can form directly in the disc wind, likely relatively near the base, and its shape directly traces the kinematics of the wind \citep[e.g.][]{matthews2023}. The other two scenarios both require that the mass that ends up producing the lines is ultimately sourced from somewhere other than the disc. In scenario B, $\epsilon_w > 1$, the disc wind acts primarily as an agent of energy- and momentum-transfer, while the mass is sourced from the immediate surroundings (e.g. the BLR) of the quasar. Finally, the quasar radiation pressure can act directly on BLR or circumnuclear gas and accelerate it. In both scenarios B and C, the wind efficiency will be limited by the physics of the quasar to gas coupling and whether it is energy- or momentum-limited, which bears a close relation to discussions of energy- and momentum-driven quasar feedback \citep[e.g.][]{costa_feedback_2014}. Finally, it is of course possible that some combination of these three scenarios at work -- maybe the wind is clumpy, but also mass-loaded from the surroundings, or the quasar radiation pressure accelerates circumnuclear material into a clumpy outflow. 

\section{Wind driving mechanisms}
\label{sec:driving}
Winds can be driven from an accretion disc by thermal, magnetic or radiation forces, as can be seen by writing down the Lagrangian form of the momentum equation in MHD \citep[e.g.][]{proga_theory_2005,reynolds_constraints_2012}. We discuss each of these mechanisms in turn and discuss their prospects for producing the required conditions for blueshifted emission lines. More specifically, we examine two main criteria. We first seek to understand whether a smooth wind driven by the mechanism in question can work -- this is equivalent to {\em estimating the maximum mass-loss rate}. Failing that, we then ask if a given driving mechanism can drive a {\em powerful} enough wind for the blueshifted lines to form either in a clumpy disc wind directly, or in the outflowing mass swept up by the disc wind. 

\subsection{Thermal Driving}
Thermal winds, and specifically Compton-heated winds, are good candidates for driving the winds observed in X-ray binaries. \cite{begelman_compton_1983} introduce much of the analytic theory, which establishes the critical radius as the Compton radius, $R_C$. This is the radius at which the escape speed is the sound speed at the Compton temperature, and is given by $R_C \approx 10^5 R_g (T_C/(10^8~{\rm K})$. The characteristic velocities at these radii are $v_{\rm esc} (R_C) = c_s(T_C) \approx 4000~{\rm km~s}^{-1}$, which is rather similar to the maximum \civ\ blueshifts. However, a strong argument against a Compton-heated wind in this case is that the gas is heated to the Compton temperature, which is typically $\gtrsim 10^6~{\rm K}$ for reasonable quasar SEDs. On this upper branch of the thermal stability `S-curve', there are few bound electrons and the \civ\ ion fraction is extremely low. Thus, while the characteristic velocities may be about right to explain \civ\ blueshifts, the wind would have to clump and cool quite dramatically for the ionization state to drop sufficiently for effective \civline\ line formation. 

Radiation-hydrodynamic simulations of thermal winds in XRBs typically find mass-loading factors of a few \citep[e.g.][]{woods_x-rayheated_1996,higginbottom_luminosity_2019}, although these may be increased a little if the wind is thermal-radiative (i.e. the thermal wind is enhanced by radiation pressure; \citealt{proga_role_2002,done_thermal_2018,tomaru_thermal-radiative_2019,higginbottom_luminosity_2019}). These mass-loading factors can be maintained if the mass is lost from the outer disk, but, for a smooth wind, they nonetheless do not reach our desired values of $\epsilon_w$. Mass-loading can result from the sweeping up of another mass reservoir providing the wind has sufficient power. However,  the kinetic power of the wind is rather low given the modest velocities ($\beta_w \sim 0.01$), with 
\begin{equation}
\dot{E}_{w,{\rm th}} \sim  
2 \times 10^{42}~{\rm erg~s^{-1}}
\left( \frac{L_{\rm bol}}{10^{46}~{\rm erg~s^{-1}}} \right)
\frac{\dot{M}_{w}}{2\dot{M}} \, .
\end{equation}
A thermal wind therefore does not have enough power to drive a smooth outflow with sufficient mass flux to explain \civ\ blueshifts, even if this mass is swept up from the surroundings, and is also disfavoured on the grounds of ionization state and temperature. 

\subsection{Magnetic Driving}
\label{sec:mhd}
An outflow can be accelerated magneto-centrifugally by large-scale poloidal magnetic fields anchored in an accretion disc \citep{blandford_hydromagnetic_1982}. Two important cylindrical radii here are the launch radius $R_L$ and the Alfven radius $R_A$. The basic idea is that the material is centrifugally lifted off the disc and accelerated continuously until it reaches the Alfven speed. Considering the extraction of angular momentum from the disc allows the derivation of a simple relation between accretion rate and mass outflow rate given by 
\begin{equation}
\dot{M}_w = \bar{\omega}^{-2} \dot{M}_{\rm acc}
\label{eq:mdot_mhd}
\end{equation}
where $\bar{\omega}\equiv R_A/R_L$ is the length of the lever arm and is typically found to be $\sim 2-3$ from detailed MHD models \citep{pudritz_disk_2007}, immediately implying $\epsilon_w \sim 0.1$, with an upper bound of $\epsilon_w = 1$. Longer lever arms will exert greater torques and thus must transport accordingly less mass. More generally, there must be some local dissipation in the disc else the disc would not be able to turn viscous stresses into local heating and the observed continuum radiation \citep[e.g.][]{knigge_disks_1997}.

Taking equation~\ref{eq:mdot_mhd} at face value, it would seem unlikely that a smooth magneto-centrifugal wind can produce significant mass flux to explain \civ\ blueshifts. However, the $\dot{M}_{\rm acc}$ here is really a local accretion rate -- and, as in section~\ref{sec:conservation}, not necessarily the accretion rate through the inner disc that sets the radiated luminosity. In the so-called `jet emitting disc' paradigm \citep{ferreira2006,marcel_jedsad2,marcel_jedsad3}, an MHD wind rises from an accretion disc threaded by a large-scale poloidal magnetic field. In this framework, $\dot{M}_{\rm acc} (r_{\rm cyl}) \propto r_{\rm cyl}^{p_m}$, with $p_m \lesssim 1$ an MHD mass ejection index derived from a self-similar MHD solution with a given magnetisation $\mu$. Values of $p_m$ obtained in these MHD solutions are $\lesssim 0.3$ \citep{Chakravorty2016,Jacquemin2021,zimniak2024}, which from Fig.~\ref{fig:mhd} is not large enough to produce $\epsilon_w \sim 50$, instead implying $\epsilon_w \lesssim 4.6$ for $r_{\rm max} = R_{sg}$. It therefore seems unlikely that an MHD wind can drive a smooth outflow with sufficient mass flux to explain \civ\ blueshifts. However, an MHD wind that is clumpy, or that sweeps up matter from its surroundings, provides a more feasible model. 

\subsection{Radiation Driving}
Radiation fields exert a force on gas that depends on its opacity $\kappa$. The classical Eddington limit is derived by considering radiation pressure acting on free electrons such that $\kappa = \sigma_T n_e$ where $\sigma_T$ is the Thomson cross-section. Most quasars have Eddington fractions in the range $0.01-1$ \citep{Steinhardt2010,shen2013,morabito_origin_2019}, and those showing strong \civ\ blueshifts have typical Eddington fractions of $\approx 0.2-0.4$ \citep{richards_unification_2011,rankine_bal_2020,temple2023}. Thus, while the uncertainties on black hole mass and the bolometric correction are significant, it seems unlikely that quasars with strong \civ\ blueshifts are uniformly driving super-Eddington winds. A radiation-driven wind scenario therefore requires some kind of opacity enhancement or `force multiplier', albeit a relatively modest one.  

Bound-free (photoelectric) and, particularly, line transitions can dramatically enhance the opacity if ionization conditions are right; the latter leads to a class of outflows known as line-driven winds where the line transitions mediate the momentum transfer from radiation field to the flow. Line-driven winds are thought to operate in stellar winds where the theory is well-developed \citep{lucy_mass_1970,castor_radiation-driven_1975,lamers_introduction_1999,owocki_instabilities_1984}, and possibly also in cataclysmic variables (CVs). Simulations of line-driven winds in AGN and CVs have been successful in producing outflows \citep{proga1998,proga_dynamics_2000,proga2002,proga_dynamics_2004,Nomura2020,dyda2024,Dyda2025,Mosallanezhad2025}, although it the flow can easily be over-ionized \citep{sim_multidimensional_2010,higginbottom_line-driven_2014,Higginbottom2024,scepi2026}. Line-driven quasar winds are also commonly invoked in phenomenological frameworks, or to explain spectral properties at a population level \citep{richards_unification_2011,giustini,rankine_bal_2020,temple2023}. 

An alternative way to enhance the radiation force is via dust. Dust-driven winds have been invoked in models for the BLR \citep{czerny_dust_2014,Czerny2017,baskin_dust_2018,Naddaf2021,Naddaf2024}, to explain the BAL phenomenon \citep{Naddaf2023,Ishibashi2024,Gaskell2024}, or to produce a dynamical obscuring dusty `torus' \citep{krolik2007,wada2012,chan2016,williamson_3d_2019,Venanzi2020}. The physics is similar in some ways to that of line-driving, but rather than requiring bound electrons one instead requires that the dust grains have not sublimated. Of course, \civline\ is formed at temperatures much higher than the dust sublimation temperature of $\sim 1500~{\rm K}$ \citep{Barvainis1987,netzer2015}; indeed, at a basic level, the formation of a \civ\ line, whose kinematics are in turn related to the properties of line absorption in BAL quasars, might by itself point towards a line-driving association. However, a multiphase or stratified wind or failed wind is often considered \citep{Czerny2017}, and within such a model it is still possible that dust driving is important or even dominant for the BLR and for \civ\ blueshifts generally.  

In the context of our study, line and/or dust opacity provide a means for the radiation field to couple effectively to an outflow, but the momentum and energy of the radiation field is still limited, and so, depending on the number of interactions and overall dynamics of the flow, the limits derived in the previous section still apply. In the single-scattering limit, the momentum imparted to the flow is $L/c$ and the limit on the mass-loading is given by equation~\ref{eq:zeta_momentum}, so it is marginal whether a smooth wind can reach the required $\epsilon_w$. However, if many line scatters or dust absorption events occur in an optically thick wind, then the wind can become energy-conserving in the sense that its momentum is `boosted' above $L/c$ and its mass-loss rate is limited by equation~\ref{eq:energy_limit}.

\section{Discussion \& Conclusions}
\label{sec:discuss}

We have shown that the presence of \civ\ emission line blueshifts (and other blueshifted emission lines that form at comparable $\xi$) implies the existence of highly mass-loaded and/or clumpy winds in quasars. This finding comes from the requirement that sufficient mass flux is supplied through the line formation region so that material with a sufficient density and ionization state remains moving at a poloidal velocity of a few thousand km~s$^{-1}$. Our results have implications for our understanding of quasar accretion and feedback, as well as the origin of the broad-line region, but there are, naturally, still several outstanding puzzles. 

One important question to understand is exactly how the blueshifted \civ\ emission connects to the BAL phenomenon. There is a relatively clear empirical connection -- the strongest, fastest BALs are generally associated with large emission line blueshifts \citep{rankine_bal_2020,rodriguez2022} -- but how does this translate to a real physical situation?  For example, are the \civ\ emission line blueshifts produced in the base of a wind that accelerates to larger velocities and also produces BALs at certain sightlines? \citetalias{matthews2023} showed that this physical association is possible, but that does not mean it plays out in nature. The true relation between these two outflow indicators in quasars is important to constrain, since it determines whether both spectral signatures can be used to constrain the same class of quasar winds. If they can, there is then more information available about the terminal velocities and ionization structures of the winds, as well as the kinematics of the BLR more generally. 

It is not yet clear whether the winds discussed here are significant agents of quasar feedback, and in fact the answer is closely tied to the preceding point and the exact nature of the wind. Let us assume that the kinetic power of the wind $\dot{E}_k$ must reach some critical fraction of $L_{\rm Edd}$ to be an effective feedback agent. Appropriate thresholds for feedback efficiencies  are not completely clear; various studies adopt, suggest or imply $\dot{E}/L_{\rm bol} \sim 0.05$ (or, often, $\dot{E}/L_{\rm Edd} \sim 0.05$), as a threshold for relatively strong AGN feedback \citep{king_black_2003,Springel2005,di_matteo_energy_2005,Sijacki2007,Sijacki2015,costa_feedback_2014}, while \cite{hopkins_quasar_2010} suggest $\dot{E}/L_{\rm Edd} \sim 5 \times10^{-3}$ based on a `two-stage' model. We can compare this feedback efficiency with the kinetic efficiency of the wind, which is given by
\begin{equation}
    \frac{\dot{E}_k}{L_{\rm Edd}} = \epsilon_w \beta_{\infty}^2 \frac{\lambda_{\rm Edd}}{2 \eta} \, ,
\end{equation}
where $\beta_{\infty} \equiv v_\infty / c$ is the terminal velocity of the wind normalised to $c$. 
If the wind is a clumpy disc wind with $f_V\approx 50$, $\dot{M}_w \approx \dot{M}$ then, if $\lambda_{\rm Edd} \approx 0.2$ so that $\lambda_{\rm Edd} / (2\eta) \approx 1$, the energetic efficiency is simply $\beta_{\infty}^2$. If $v_\infty = 3000~{\rm km~s}^{-1}$, $\beta_{\infty} \approx 0.01$ and $\dot{E}_k/L_{\rm Edd}\sim 10^{-4}$, which doesn't get you far in a quasar feedback sense. Another way of phrasing this is that a disc wind with $\dot{M}_w \sim \dot{M}$ has not extracted much of the quasar energy budget by the time it reaches $3000~{\rm km~s}^{-1}$. 

However, if the wind accelerates to the speeds associated with the more extreme BAL outflows \citep{hamann_extreme-velocity_2013,rodriguez2022} or UFOs \citep{pounds_high-velocity_2003,reeves_massive_2003,gofford_suzaku_2013} of $\beta_{\infty} \approx 0.1$, an $\epsilon_w=1$ wind has a kinetic luminosity that reaches $5\%$ for $\lambda_{\rm Edd} \approx 0.5$. For more modest wind speeds of $10,000~{\rm km~s}^{-1}$ then $\dot{E}_k/L_{\rm Edd} \sim 3\times10^{-3}$ for the same parameters. Smoother, more mass-loaded winds are in principle more effective feedback agents (because $\dot{E}_k \propto \dot{M}_w \propto \epsilon_w f_V$), but it seems unrealistic that a wind first sweeps up material before accelerating to BAL-like velocities. Thus, while it is not in question that the winds considered here are dynamically important in general -- especially  for the quasar disc and immediate surroundings -- it is far from certain what the real power of the winds are and whether they can be an effective agent of AGN feedback on galactic scales. 

We have discussed the potential requirement for clumping, but have not yet provided a mechanism for producing it. Clumping in winds can be driven by a variety of processes, including dynamic thermal instability \citep{Balbus1986,Dannen2020,Waters2022} or the line deshadowing instability \citep[LDI;][]{macgregor_radiative_1979,Carlberg1980,owocki_instabilities_1984,owocki_time-dependent_1988,Driessen2019}. These processes are often microscopic (in the sense that they depend on quite localised conditions), but larger-scale mechanisms, such as dynamic shielding effects within a line-driven wind can also lead to clumps developing. Furthermore, clumping could be caused by the time variability of the radiation field \citep{Waters2016}, the interaction between the wind and the disc \citep{Waters2021}, the interaction between the wind and surrounding gas, or more generally due to the action of Kelvin-Helmholtz and Rayleigh-Taylor instabilities. 

Various authors have suggested that the broad-line region is associated with a clumpy and/or turbulent outflow \citep{emmering_magnetic_1992,Baldwin1996,matthews_testing_2016,elvis_quasar_2017,waters_agn_2019}, and it is well-known that clumping is important in stellar winds \citep[e.g.][]{Pauldrach1986,moffat_clumping_1994,lamers_introduction_1999,lepine_direct_2008,Driessen2019} where the LDI is thought to operate. Indeed, it is notable that for $\epsilon_w \sim 1$ we require $f_V^{-1} \sim 50$, a clumping factor that is comparable to those predicted theoretically \citep[e.g.][]{Driessen2019} and inferred empirically \citep{crowther2002,bouret2003,hillier2003,Najarro2011,Hawcroft2024} in line-driven stellar winds. This correspondence between clumping factors could be suggestive of a line-driven mechanism at work in quasars. Indeed, clumping has also been proposed as a solution to the over-ionization of line-driven disc winds \citep{Mosallanezhad2025}. However, clumping physics is complex and the relative importance of each mechanism is very specific to the local plasma conditions, motivating further hydrodynamic simulations of quasar disc winds that fully treat the relevant radiative processes and thermodynamics.

\section*{Data Availability}
The simulations in section~\ref{sec:sirocco} were run using \sirocco\ version 1.1. \sirocco\ is hosted on \href{https://github.com/sirocco-rt/sirocco}{GitHub}, with associated documentation on \href{https://sirocco-rt.readthedocs.io/en/latest/}{ReadTheDocs}. Data is available on reasonable request. 

\section*{Acknowledgements}
I thank the anonymous referee for a helpful and constructive report. This paper is dedicated to the memory of Paul Hewett, who taught me an awful lot about quasars -- but only a tiny fraction of what he knew about them. Paul was a fine scientist and a kind and supportive mentor to many astronomers in the UK and beyond; he will be missed immensely. I would like to thank Christian Knigge,  Matthew Temple, Knox Long, the \textsc{Sirocco} collaboration, and many others for helpful discussions and general encouragement. I acknowledge funding from a Royal Society URF (URF\textbackslash R1\textbackslash 21062). I acknowledge the use of ChatGPT for some conceptual feedback, proof-reading and limited code/plotting assistance. The final content remains the sole responsibility of the author. I gratefully acknowledge the use of the following software packages: astropy \citep{astropy-collaboration13,astropy-collaboration18}, {\sc Cloudy} \citep{ferland17}, matplotlib \citep{hunter07}, GNU Science Library \citep{Galassi2018_gsl}. 

\bibliographystyle{mnras}
\bibliography{extracted} 

@article{fender_towards_2004,
  author =        {Fender, R. P. and Belloni, T. M. and Gallo, E.},
  journal =       {\mnras},
  month =         dec,
  pages =         {1105--1118},
  title =         {Towards a unified model for black hole {X}-ray binary
                   jets},
  volume =        {355},
  year =          {2004},
  doi =           {10.1111/j.1365-2966.2004.08384.x},
}

@ARTICLE{marcel_jedsad2,
       author = {{Marcel}, G. and {Ferreira}, J. and {Petrucci}, P. -O. and {Henri}, G. and {Belmont}, R. and {Clavel}, M. and {Malzac}, J. and {Coriat}, M. and {Corbel}, S. and {Rodriguez}, J. and {Loh}, A. and {Chakravorty}, S. and {Drappeau}, S.},
        title = "{A unified accretion-ejection paradigm for black hole X-ray binaries. II. Observational signatures of jet-emitting disks}",
      journal = {\aap},
         year = 2018,
        month = jul,
       volume = {615},
          eid = {A57},
        pages = {A57},
          doi = {10.1051/0004-6361/201732069},
archivePrefix = {arXiv},
       eprint = {1803.04335},
 primaryClass = {astro-ph.HE},
       adsurl = {https://ui.adsabs.harvard.edu/abs/2018A&A...615A..57M}
}

@ARTICLE{marcel_jedsad3,
       author = {{Marcel}, G. and {Ferreira}, J. and {Petrucci}, P.-O. and {Belmont}, R. and {Malzac}, J. and {Clavel}, M. and {Henri}, G. and {Coriat}, M. and {Corbel}, S. and {Rodriguez}, J. and {Loh}, A. and {Chakravorty}, S.},
        title = "{A unified accretion-ejection paradigm for black hole X-ray binaries. III. Spectral signatures of hybrid disk configurations}",
      journal = {\aap},
         year = 2018,
        month = sep,
       volume = {617},
          eid = {A46},
        pages = {A46},
          doi = {10.1051/0004-6361/201833124},
archivePrefix = {arXiv},
       eprint = {1805.12407},
 primaryClass = {astro-ph.HE},
       adsurl = {https://ui.adsabs.harvard.edu/abs/2018A&A...617A..46M}
}

@ARTICLE{ferreira2006,
       author = {{Ferreira}, J. and {Petrucci}, P.-O. and {Henri}, G. and {Saug{\'e}}, L. and {Pelletier}, G.},
        title = "{A unified accretion-ejection paradigm for black hole X-ray binaries. I. The dynamical constituents}",
      journal = {\aap},
         year = 2006,
        month = mar,
       volume = {447},
       number = {3},
        pages = {813-825},
          doi = {10.1051/0004-6361:20052689},
archivePrefix = {arXiv},
       eprint = {astro-ph/0511123},
 primaryClass = {astro-ph},
       adsurl = {https://ui.adsabs.harvard.edu/abs/2006A&A...447..813F}
}

@ARTICLE{Collin-Souffrin1990,
       author = {{Collin-Souffrin}, S. and {Dumont}, A.~M.},
        title = "{Line and continuum emission from the outer regions of accretion discs in active galactic nuclei. II. Radial structure of the disc.}",
      journal = {\aap},
         year = 1990,
        month = mar,
       volume = {229},
        pages = {292-301},
       adsurl = {https://ui.adsabs.harvard.edu/abs/1990A&A...229..292C}
}

@ARTICLE{Jacquemin2021,
       author = {{Jacquemin-Ide}, J. and {Lesur}, G. and {Ferreira}, J.},
        title = "{Magnetic outflows from turbulent accretion disks. I. Vertical structure and secular evolution}",
      journal = {\aap},
         year = 2021,
        month = mar,
       volume = {647},
          eid = {A192},
        pages = {A192},
          doi = {10.1051/0004-6361/202039322},
archivePrefix = {arXiv},
       eprint = {2011.14782},
 primaryClass = {astro-ph.HE},
       adsurl = {https://ui.adsabs.harvard.edu/abs/2021A&A...647A.192J}
}

@ARTICLE{zimniak2024,
       author = {{Zimniak}, N. and {Ferreira}, J. and {Jacquemin-Ide}, J.},
        title = "{Influence of the turbulent magnetic pressure on isothermal jet emitting disks}",
      journal = {\aap},
         year = 2024,
        month = dec,
       volume = {692},
          eid = {A99},
        pages = {A99},
          doi = {10.1051/0004-6361/202450501},
archivePrefix = {arXiv},
       eprint = {2412.06999},
 primaryClass = {astro-ph.HE},
       adsurl = {https://ui.adsabs.harvard.edu/abs/2024A&A...692A..99Z}
}

@article{ponti_ubiquitous_2012,
  author =        {Ponti, G. and Fender, R. P. and Begelman, M. C. and
                   Dunn, R. J. H. and Neilsen, J. and Coriat, M.},
  journal =       {\mnras},
  month =         may,
  pages =         {L11},
  title =         {Ubiquitous equatorial accretion disc winds in black
                   hole soft states},
  volume =        {422},
  year =          {2012},
  doi =           {10.1111/j.1745-3933.2012.01224.x},
}

@article{rankine_bal_2020,
  author =        {Rankine, Amy L. and Hewett, Paul C. and
                   Banerji, Manda and Richards, Gordon T.},
  journal =       {\mnras},
  month =         mar,
  pages =         {4553--4575},
  title =         {{BAL} and non-{BAL} quasars: continuum, emission, and
                   absorption properties establish a common parent
                   sample},
  volume =        {492},
  year =          {2020},
  doi =           {10.1093/mnras/staa130},
  issn =          {0035-8711},
  url =           {https://ui.adsabs.harvard.edu/abs/2020MNRAS.492.4553R},
}

@article{temple2023,
  author =        {{Temple}, Matthew J. and {Matthews}, James H. and
                   {Hewett}, Paul C. and {Rankine}, Amy L. and
                   {Richards}, Gordon T. and {Banerji}, Manda and
                   {Ferland}, Gary J. and {Knigge}, Christian and
                   {Stepney}, Matthew},
  journal =       {\mnras},
  month =         jul,
  number =        {1},
  pages =         {646-666},
  title =         {{Testing AGN outflow and accretion models with C IV
                   and He II emission line demographics in z
                   {\ensuremath{\approx}} 2 quasars}},
  volume =        {523},
  year =          {2023},
  doi =           {10.1093/mnras/stad1448},
}

@article{Parra2024,
  author =        {{Parra}, M. and {Petrucci}, P.-O. and {Bianchi}, S. and
                   {Gianolli}, V.~E. and {Ursini}, F. and {Ponti}, G.},
  journal =       {\aap},
  month =         jan,
  pages =         {A49},
  title =         {{The current state of disk wind observations in
                   BHLMXBs through X-ray absorption lines in the iron
                   band}},
  volume =        {681},
  year =          {2024},
  doi =           {10.1051/0004-6361/202346920},
  eid =           {A49},
}

@article{jackson2026,
  author =        {{Jackson}, Charlotte L. and {Matthews}, James H. and
                   {Whittam}, Imogen H. and {Jarvis}, Matt J. and
                   {Temple}, Matthew J. and {Rankine}, Amy L. and
                   {Hewett}, Paul C.},
  journal =       {\mnras},
  month =         mar,
  number =        {3},
  pages =         {stag065},
  title =         {{Exploring the quasar disc─wind─jet connection
                   with LoTSS and SDSS}},
  volume =        {546},
  year =          {2026},
  doi =           {10.1093/mnras/stag065},
  eid =           {stag065},
}

@article{morganti_many_2017,
  author =        {Morganti, Raffaella},
  journal =       {Frontiers in Astronomy and Space Sciences},
  month =         nov,
  pages =         {42},
  title =         {The many routes to {AGN} feedback},
  volume =        {4},
  year =          {2017},
  doi =           {10.3389/fspas.2017.00042},
  url =           {http://adsabs.harvard.edu/abs/2017FrASS...4...42M},
}

@article{silk_quasars_1998,
  author =        {Silk, Joseph and Rees, Martin J.},
  journal =       {\aap},
  month =         mar,
  pages =         {L1--L4},
  title =         {Quasars and galaxy formation},
  volume =        {331},
  year =          {1998},
  issn =          {0004-6361},
  url =           {http://adsabs.harvard.edu/abs/1998A%26A...331L...1S},
}

@article{king_black_2003,
  author =        {King, A.},
  journal =       {\apjl},
  month =         oct,
  pages =         {L27--L29},
  title =         {Black {Holes}, {Galaxy} {Formation}, and the
                   {M}$_{\textrm{{BH}}}$-σ {Relation}},
  volume =        {596},
  year =          {2003},
  doi =           {10.1086/379143},
}

@article{di_matteo_energy_2005,
  author =        {Di Matteo, Tiziana and Springel, Volker and
                   Hernquist, Lars},
  journal =       {Nature},
  month =         feb,
  note =          {ADS Bibcode: 2005Natur.433..604D},
  pages =         {604--607},
  title =         {Energy input from quasars regulates the growth and
                   activity of black holes and their host galaxies},
  volume =        {433},
  year =          {2005},
  doi =           {10.1038/nature03335},
  issn =          {0028-0836},
  url =           {https://ui.adsabs.harvard.edu/abs/2005Natur.433..604D},
}

@article{hopkins_quasar_2010,
  author =        {Hopkins, P. F. and Elvis, M.},
  journal =       {\mnras},
  month =         jan,
  pages =         {7--14},
  title =         {Quasar feedback: more bang for your buck},
  volume =        {401},
  year =          {2010},
  doi =           {10.1111/j.1365-2966.2009.15643.x},
}

@article{costa_feedback_2014,
  author =        {Costa, Tiago and Sijacki, Debora and
                   Haehnelt, Martin G.},
  journal =       {\mnras},
  month =         nov,
  pages =         {2355--2376},
  title =         {Feedback from active galactic nuclei: energy- versus
                   momentum-driving},
  volume =        {444},
  year =          {2014},
  doi =           {10.1093/mnras/stu1632},
  issn =          {0035-8711},
  url =           {http://adsabs.harvard.edu/abs/2014MNRAS.444.2355C},
}

@article{harrison_agn_2018,
  author =        {Harrison, C. M. and Costa, T. and Tadhunter, C. N. and
                   Flütsch, A. and Kakkad, D. and Perna, M. and
                   Vietri, G.},
  journal =       {Nature Astronomy},
  month =         mar,
  note =          {arXiv: 1802.10306},
  number =        {3},
  pages =         {198--205},
  title =         {{AGN} outflows and feedback twenty years on},
  volume =        {2},
  year =          {2018},
  doi =           {10.1038/s41550-018-0403-6},
  issn =          {2397-3366},
  url =           {http://arxiv.org/abs/1802.10306},
}

@article{laha_ionized_2021,
  author =        {Laha, Sibasish and Reynolds, Christopher S. and
                   Reeves, James and Kriss, Gerard and Guainazzi, Matteo and
                   Smith, Randall and Veilleux, Sylvain and
                   Proga, Daniel},
  journal =       {Nature Astronomy},
  month =         jan,
  note =          {ADS Bibcode: 2021NatAs...5...13L},
  pages =         {13--24},
  title =         {Ionized outflows from active galactic nuclei as the
                   essential elements of feedback},
  volume =        {5},
  year =          {2021},
  doi =           {10.1038/s41550-020-01255-2},
  issn =          {2397-3366},
  url =           {https://ui.adsabs.harvard.edu/abs/2021NatAs...5...13L},
}

@article{heap_iue_1978,
  author =        {Heap, S. R. and Boggess, A. and Holm, A. and
                   Klinglesmith, D. A. and Sparks, W. and West, D. and
                   Wu, C. C. and Boksenberg, A. and Willis, A. and
                   Wilson, R. and Macchetto, F. and Selvelli, P. O. and
                   Stickland, D. and Greenstein, J. L. and
                   Hutchings, J. B. and Underhill, A. B. and Viotti, R. and
                   Whelan, J. A. J.},
  journal =       {Nature},
  month =         oct,
  pages =         {385--388},
  title =         {{IUE} observations of hot stars - {HZ43}, {BD} +75
                   deg 325, {NGC} 6826, {SS} {Cygni}, {Eta} {Carinae}},
  volume =        {275},
  year =          {1978},
  doi =           {10.1038/275385a0},
  issn =          {0028-0836},
  url =           {http://adsabs.harvard.edu/abs/1978Natur.275..385H},
}

@article{greenstein_rw_1982,
  author =        {Greenstein, J. L. and Oke, J. B.},
  journal =       {\apj},
  month =         jul,
  pages =         {209--216},
  title =         {{RW} {Sextantis}, a disk with a hot, high-velocity
                   wind},
  volume =        {258},
  year =          {1982},
  doi =           {10.1086/160069},
}

@article{cordova_high-velocity_1982,
  author =        {Cordova, F. A. and Mason, K. O.},
  journal =       {\apj},
  month =         sep,
  pages =         {716--721},
  title =         {High-velocity winds from a dwarf nova during
                   outburst},
  volume =        {260},
  year =          {1982},
  doi =           {10.1086/160291},
}

@article{kafka_detecting_2004,
  author =        {Kafka, S. and Honeycutt, R. K.},
  journal =       {\aj},
  month =         nov,
  pages =         {2420--2429},
  title =         {Detecting {Outflows} from {Cataclysmic} {Variables}
                   in the {Optical}},
  volume =        {128},
  year =          {2004},
  doi =           {10.1086/424618},
  issn =          {0004-6256},
  url =           {http://adsabs.harvard.edu/abs/2004AJ....128.2420K},
}

@article{Ioannou2003,
  author =        {{Ioannou}, Z. and {van Zyl}, L. and {Naylor}, T. and
                   {Charles}, P.~A. and {Margon}, B. and
                   {Koch-Miramond}, L. and {Ilovaisky}, S.},
  journal =       {\aap},
  month =         feb,
  pages =         {211-218},
  title =         {{Understanding the LMXB X2127+119 in M 15. II. The UV
                   data}},
  volume =        {399},
  year =          {2003},
  doi =           {10.1051/0004-6361:20021578},
}

@article{munoz-darias_regulation_2016,
  author =        {Muñoz-Darias, T. and Casares, J. and
                   Mata Sánchez, D. and Fender, R. P. and
                   Armas Padilla, M. and Linares, M. and Ponti, G. and
                   Charles, P. A. and Mooley, K. P. and Rodriguez, J.},
  journal =       {\nat},
  month =         may,
  title =         {Regulation of black-hole accretion by a disk wind
                   during a violent outburst of {V404} {Cygni}},
  year =          {2016},
  doi =           {10.1038/nature17446},
}

@article{fijma2023,
  author =        {{Fijma}, S. and {Castro Segura}, N. and
                   {Degenaar}, N. and {Knigge}, C. and
                   {Higginbottom}, N. and
                   {Hern{\'a}ndez Santisteban}, J.~V. and
                   {Maccarone}, T.~J.},
  journal =       {\mnras},
  month =         nov,
  number =        {1},
  pages =         {L149-L154},
  title =         {{A transient ultraviolet outflow in the short-period
                   X-ray binary UW CrB}},
  volume =        {526},
  year =          {2023},
  doi =           {10.1093/mnrasl/slad125},
}

@article{castro2022,
  author =        {{Castro Segura}, N. and {Knigge}, C. and
                   {Long}, K.~S. and {Altamirano}, D. and
                   {Armas Padilla}, M. and {Bailyn}, C. and
                   {Buckley}, D.~A.~H. and {Buisson}, D.~J.~K. and
                   {Casares}, J. and {Charles}, P. and {Combi}, J.~A. and
                   {C{\'u}neo}, V.~A. and {Degenaar}, N.~D. and
                   {del Palacio}, S. and {D{\'\i}az Trigo}, M. and
                   {Fender}, R. and {Gandhi}, P. and {Georganti}, M. and
                   {Guti{\'e}rrez}, C. and
                   {Hernandez Santisteban}, J.~V. and
                   {Jim{\'e}nez-Ibarra}, F. and {Matthews}, J. and
                   {M{\'e}ndez}, M. and {Middleton}, M. and
                   {Mu{\~n}oz-Darias}, T. and {{\"O}zbey Arabac{\i}}, M. and
                   {Pahari}, M. and {Rhodes}, L. and {Russell}, T.~D. and
                   {Scaringi}, S. and {van den Eijnden}, J. and
                   {Vasilopoulos}, G. and {Vincentelli}, F.~M. and
                   {Wiseman}, P.},
  journal =       {\nat},
  month =         mar,
  number =        {7899},
  pages =         {52-57},
  title =         {{A persistent ultraviolet outflow from an accreting
                   neutron star binary transient}},
  volume =        {603},
  year =          {2022},
  doi =           {10.1038/s41586-021-04324-2},
}

@article{Castro2026,
  author =        {{Castro Segura}, N. and {Solomons}, K. and
                   {Corral-Santana}, J.~M. and {Knigge}, C. and
                   {Charles}, P.~A. and {Brigitte}, M. and {Fijma}, S. and
                   {Diaz-Trigo}, M. and {G{\'u}rpide}, A. and
                   {Buckley}, D.~A.~H. and {Carotenuto}, F. and
                   {Castro-Tirado}, A.~J. and {Coppejans}, D.~L. and
                   {Georganti}, M. and {Hughes}, A. and {Long}, K.~S. and
                   {Matthews}, J. and {Monageng}, I. and {Pelisoli}, I. and
                   {Russell}, T.~D. and {Steeghs}, D. and {Svoboda}, J. and
                   {Tetarenko}, A.~J. and {Vincentelli}, F.~M. and
                   {Wallis}, A.~G.~W.},
  journal =       {arXiv e-prints},
  month =         mar,
  pages =         {arXiv:2603.17023},
  title =         {{Optical outburst evolution of the transient black
                   hole X-ray binary Swift J1727.8-1613: Disc response
                   to jet ejections and late-outburst emergence of
                   powerful disc winds}},
  year =          {2026},
  doi =           {10.48550/arXiv.2603.17023},
  eid =           {arXiv:2603.17023},
}

@article{bastian1985,
  author =        {{Bastian}, U. and {Mundt}, R.},
  journal =       {\aap},
  month =         mar,
  pages =         {57-63},
  title =         {{FU Orionis Star Winds}},
  volume =        {144},
  year =          {1985},
}

@article{hartmann1996,
  author =        {{Hartmann}, Lee and {Kenyon}, Scott J.},
  journal =       {\araa},
  month =         jan,
  pages =         {207-240},
  title =         {{The FU Orionis Phenomenon}},
  volume =        {34},
  year =          {1996},
  doi =           {10.1146/annurev.astro.34.1.207},
}

@article{milliner_disc_2019,
  author =        {Milliner, Kelly and Matthews, James H. and
                   Long, Knox S. and Hartmann, Lee},
  journal =       {\mnras},
  month =         feb,
  pages =         {1663--1673},
  title =         {Disc wind models for {FU} {Ori} objects},
  volume =        {483},
  year =          {2019},
  doi =           {10.1093/mnras/sty3197},
  issn =          {0035-8711},
  url =           {https://ui.adsabs.harvard.edu/abs/2019MNRAS.483.1663M},
}

@article{hewett2003,
  author =        {{Hewett}, Paul C. and {Foltz}, Craig B.},
  journal =       {\aj},
  month =         apr,
  number =        {4},
  pages =         {1784-1794},
  title =         {{The Frequency and Radio Properties of Broad
                   Absorption Line Quasars}},
  volume =        {125},
  year =          {2003},
  doi =           {10.1086/368392},
}

@article{weymann_comparisons_1991,
  author =        {Weymann, R. J. and Morris, S. L. and Foltz, C. B. and
                   Hewett, P. C.},
  journal =       {\apj},
  month =         may,
  pages =         {23--53},
  title =         {Comparisons of the emission-line and continuum
                   properties of broad absorption line and normal
                   quasi-stellar objects},
  volume =        {373},
  year =          {1991},
  doi =           {10.1086/170020},
}

@article{pounds_high-velocity_2003,
  author =        {Pounds, K. A. and Reeves, J. N. and King, A. R. and
                   Page, K. L. and O'Brien, P. T. and Turner, M. J. L.},
  journal =       {\mnras},
  month =         nov,
  number =        {3},
  pages =         {705--713},
  title =         {A high-velocity ionized outflow and {XUV} photosphere
                   in the narrow emission line quasar {PG1211}+143},
  volume =        {345},
  year =          {2003},
  doi =           {10.1046/j.1365-8711.2003.07006.x},
  issn =          {0035-8711},
  language =      {en},
  url =           {https://academic.oup.com/mnras/article/345/3/705/972787},
}

@article{gofford_suzaku_2013,
  author =        {Gofford, Jason and Reeves, James N. and
                   Tombesi, Francesco and Braito, Valentina and
                   Turner, T. Jane and Miller, Lance and Cappi, Massimo},
  journal =       {\mnras},
  month =         mar,
  number =        {1},
  pages =         {60--80},
  title =         {The {Suzaku} view of highly ionized outflows in {AGN}
                   – {I}. {Statistical} detection and global absorber
                   properties},
  volume =        {430},
  year =          {2013},
  doi =           {10.1093/mnras/sts481},
  issn =          {0035-8711},
  language =      {en},
  url =           {https://academic.oup.com/mnras/article/430/1/60/983995},
}

@article{tombesi_unification_2013,
  author =        {Tombesi, F. and Cappi, M. and Reeves, J. N. and
                   Nemmen, R. S. and Braito, V. and Gaspari, M. and
                   Reynolds, C. S.},
  journal =       {\mnras},
  month =         apr,
  pages =         {1102--1117},
  title =         {Unification of {X}-ray winds in {Seyfert} galaxies:
                   from ultra-fast outflows to warm absorbers},
  volume =        {430},
  year =          {2013},
  doi =           {10.1093/mnras/sts692},
  url =           {https://ui.adsabs.harvard.edu/abs/2013MNRAS.430.1102T},
}

@article{laha_warm_2014,
  author =        {Laha, Sibasish and Guainazzi, Matteo and
                   Dewangan, Gulab C. and Chakravorty, Susmita and
                   Kembhavi, Ajit K.},
  journal =       {\mnras},
  month =         jul,
  number =        {3},
  pages =         {2613--2643},
  title =         {Warm absorbers in {X}-rays ({WAX}), a comprehensive
                   high-resolution grating spectral study of a sample of
                   {Seyfert} galaxies – {I}. {A} global view and
                   frequency of occurrence of warm absorbers.},
  volume =        {441},
  year =          {2014},
  doi =           {10.1093/mnras/stu669},
  issn =          {0035-8711},
  language =      {en},
  url =           {https://academic.oup.com/mnras/article/441/3/2613/1112542},
}

@article{gaskell_redshift_1982,
  author =        {Gaskell, C. M.},
  journal =       {\apj},
  month =         dec,
  pages =         {79--86},
  title =         {A redshift difference between high and low ionization
                   emission-line regions in {QSO}'s-evidence for radial
                   motions.},
  volume =        {263},
  year =          {1982},
  doi =           {10.1086/160481},
  issn =          {0004-637X},
  url =           {https://ui.adsabs.harvard.edu/abs/1982ApJ...263...79G},
}

@article{wilkes_studies_1984,
  author =        {Wilkes, B. J.},
  journal =       {\mnras},
  month =         mar,
  note =          {ADS Bibcode: 1984MNRAS.207...73W},
  pages =         {73--98},
  title =         {Studies of broad emission line profiles in {QSOs} -
                   {I}. {Observed}, high-resolution profiles.},
  volume =        {207},
  year =          {1984},
  doi =           {10.1093/mnras/207.1.73},
  issn =          {0035-8711},
  url =           {https://ui.adsabs.harvard.edu/abs/1984MNRAS.207...73W},
}

@article{richards_unification_2011,
  author =        {Richards, Gordon T. and Kruczek, Nicholas E. and
                   Gallagher, S. C. and Hall, Patrick B. and
                   Hewett, Paul C. and Leighly, Karen M. and
                   Deo, Rajesh P. and Kratzer, Rachael M. and Shen, Yue},
  journal =       {\aj},
  month =         may,
  pages =         {167},
  title =         {Unification of {Luminous} {Type} 1 {Quasars} through
                   {C} {IV} {Emission}},
  volume =        {141},
  year =          {2011},
  doi =           {10.1088/0004-6256/141/5/167},
  issn =          {0004-6256},
  url =           {http://adsabs.harvard.edu/abs/2011AJ....141..167R},
}

@article{rankine_placing_2021,
  author =        {Rankine, Amy L. and Matthews, James H. and
                   Hewett, Paul C. and Banerji, Manda and
                   Morabito, Leah K. and Richards, Gordon T.},
  journal =       {\mnras},
  month =         apr,
  pages =         {4154--4169},
  title =         {Placing {LOFAR}-detected quasars in {C} {IV} emission
                   space: implications for winds, jets and star
                   formation},
  volume =        {502},
  year =          {2021},
  doi =           {10.1093/mnras/stab302},
  issn =          {0035-8711},
  url =           {https://ui.adsabs.harvard.edu/abs/2021MNRAS.502.4154R},
}

@article{richards_probing_2021,
  author =        {Richards, Gordon T. and McCaffrey, Trevor V. and
                   Kimball, Amy and Rankine, Amy L. and
                   Matthews, James H. and Hewett, Paul C. and
                   Rivera, Angelica B.},
  journal =       {\aj},
  month =         dec,
  pages =         {270},
  title =         {Probing the {Wind} {Component} of {Radio} {Emission}
                   in {Luminous} {High}-redshift {Quasars}},
  volume =        {162},
  year =          {2021},
  doi =           {10.3847/1538-3881/ac283b},
  issn =          {0004-6256},
  url =           {https://ui.adsabs.harvard.edu/abs/2021AJ....162..270R},
}

@article{petley_connecting_2022,
  author =        {Petley, J. W. and Morabito, L. K. and
                   Alexander, D. M. and Rankine, A. L. and
                   Fawcett, V. A. and Rosario, D. J. and Matthews, J. H. and
                   Shimwell, T. M. and Drabent, A.},
  journal =       {\mnras},
  month =         oct,
  note =          {ADS Bibcode: 2022MNRAS.515.5159P},
  pages =         {5159--5174},
  title =         {Connecting radio emission to {AGN} wind properties
                   with broad absorption line quasars},
  volume =        {515},
  year =          {2022},
  doi =           {10.1093/mnras/stac2067},
  issn =          {0035-8711},
  url =           {https://ui.adsabs.harvard.edu/abs/2022MNRAS.515.5159P},
}

@article{richards_broad_2002,
  author =        {Richards, Gordon T. and Vanden Berk, Daniel E. and
                   Reichard, Timothy A. and Hall, Patrick B. and
                   Schneider, Donald P. and SubbaRao, Mark and
                   Thakar, Anirudda R. and York, Donald G.},
  journal =       {\aj},
  month =         jul,
  pages =         {1--17},
  title =         {Broad {Emission}-{Line} {Shifts} in {Quasars}: {An}
                   {Orientation} {Measure} for {Radio}-{Quiet}
                   {Quasars}?},
  volume =        {124},
  year =          {2002},
  doi =           {10.1086/341167},
  issn =          {0004-6256},
  url =           {https://ui.adsabs.harvard.edu/abs/2002AJ....124....1R},
}

@article{yong_kinematics_2017,
  author =        {Yong, Suk Yee and Webster, Rachel L. and
                   King, Anthea L. and Bate, Nicholas F. and
                   O’Dowd, Matthew J. and Labrie, Kathleen},
  journal =       {Publications of the Astronomical Society of
                   Australia},
  title =         {The {Kinematics} of {Quasar} {Broad} {Emission}
                   {Line} {Regions} {Using} a {Disk}-{Wind} {Model}},
  volume =        {34},
  year =          {2017},
  doi =           {10.1017/pasa.2017.37},
  issn =          {1323-3580, 1448-6083},
  language =      {en},
  url =           {https://www.cambridge.org/core/journals/publications-of-the-
                  astronomical-society-of-australia/article/kinematics-of-
                  quasar-broad-emission-line-regions-using-a-diskwind-model/
                  00AB7BB064A4FC661716C944DC6F6113},
}

@article{matthews2023,
  author =        {{Matthews}, James H. and {Strong-Wright}, Jago and
                   {Knigge}, Christian and {Hewett}, Paul and
                   {Temple}, Matthew J. and {Long}, Knox S. and
                   {Rankine}, Amy L. and {Stepney}, Matthew and
                   {Banerji}, Manda and {Richards}, Gordon T.},
  journal =       {\mnras},
  month =         dec,
  number =        {3},
  pages =         {3967-3986},
  title =         {{A disc wind model for blueshifts in quasar broad
                   emission lines}},
  volume =        {526},
  year =          {2023},
  doi =           {10.1093/mnras/stad2895},
}

@article{gaskell_case_2016,
  author =        {Gaskell, C. Martin and Goosmann, René W.},
  journal =       {Astrophysics and Space Science},
  month =         feb,
  note =          {ADS Bibcode: 2016Ap\&SS.361...67G},
  pages =         {67},
  title =         {The case for inflow of the broad-line region of
                   active galactic nuclei},
  volume =        {361},
  year =          {2016},
  doi =           {10.1007/s10509-015-2648-1},
  issn =          {0004-640X},
  url =           {https://ui.adsabs.harvard.edu/abs/2016Ap&SS.361...67G},
}

@article{giustini,
  author =        {{Giustini}, Margherita and {Proga}, Daniel},
  journal =       {\aap},
  month =         oct,
  pages =         {A94},
  title =         {{A global view of the inner accretion and ejection
                   flow around super massive black holes.
                   Radiation-driven accretion disk winds in a physical
                   context}},
  volume =        {630},
  year =          {2019},
  doi =           {10.1051/0004-6361/201833810},
  eid =           {A94},
}

@article{reynolds_constraints_2012,
  author =        {Reynolds, Christopher S.},
  journal =       {\apj},
  month =         nov,
  note =          {ADS Bibcode: 2012ApJ...759L..15R},
  pages =         {L15},
  title =         {Constraints on {Compton}-thick {Winds} from {Black}
                   {Hole} {Accretion} {Disks}: {Can} {We} {See} the
                   {Inner} {Disk}?},
  volume =        {759},
  year =          {2012},
  doi =           {10.1088/2041-8205/759/1/L15},
  issn =          {0004-637X},
  url =           {https://ui.adsabs.harvard.edu/abs/2012ApJ...759L..15R},
}

@article{richards_sloan_2006,
  author =        {Richards, Gordon T. and Strauss, Michael A. and
                   Fan, Xiaohui and Hall, Patrick B. and
                   Jester, Sebastian and Schneider, Donald P. and
                   Vanden Berk, Daniel E. and Stoughton, Chris and
                   Anderson, Scott F. and Brunner, Robert J. and
                   Gray, Jim and Gunn, James E. and Ivezić, Željko and
                   Kirkland, Margaret K. and Knapp, G. R. and
                   Loveday, Jon and Meiksin, Avery and Pope, Adrian and
                   Szalay, Alexander S. and Thakar, Anirudda R. and
                   Yanny, Brian and York, Donald G. and Barentine, J. C. and
                   Brewington, Howard J. and Brinkmann, J. and
                   Fukugita, Masataka and Harvanek, Michael and
                   Kent, Stephen M. and Kleinman, S. J. and
                   Krzesiński, Jurek and Long, Daniel C. and
                   Lupton, Robert H. and Nash, Thomas and
                   Neilsen, Eric H. and Nitta, Atsuko and
                   Schlegel, David J. and Snedden, Stephanie A.},
  journal =       {\aj},
  month =         jun,
  number =        {6},
  pages =         {2766},
  title =         {The {Sloan} {Digital} {Sky} {Survey} {Quasar}
                   {Survey}: {Quasar} {Luminosity} {Function} from
                   {Data} {Release} 3},
  volume =        {131},
  year =          {2006},
  doi =           {10.1086/503559},
  language =      {en},
  url =           {https://ui.adsabs.harvard.edu/#abs/2006AJ....131.2766R/
                  abstract},
}

@article{matthews_stratified_2020,
  author =        {Matthews, James H. and Knigge, Christian and
                   Higginbottom, Nick and Long, Knox S. and
                   Sim, Stuart A. and Mangham, Samuel W. and
                   Parkinson, Edward J. and Hewitt, Henrietta A.},
  journal =       {\mnras},
  month =         mar,
  pages =         {5540--5560},
  title =         {Stratified disc wind models for the {AGN} broad-line
                   region: ultraviolet, optical, and {X}-ray properties},
  volume =        {492},
  year =          {2020},
  doi =           {10.1093/mnras/staa136},
  issn =          {0035-8711},
  url =           {https://ui.adsabs.harvard.edu/abs/2020MNRAS.492.5540M},
}

@article{kubota,
  author =        {{Kubota}, Aya and {Done}, Chris},
  journal =       {\mnras},
  month =         oct,
  number =        {1},
  pages =         {1247-1262},
  title =         {{A physical model of the broad-band continuum of AGN
                   and its implications for the UV/X relation and
                   optical variability}},
  volume =        {480},
  year =          {2018},
  doi =           {10.1093/mnras/sty1890},
}

@article{tarter_interaction_1969,
  author =        {Tarter, C. Bruce and Tucker, Wallace H. and
                   Salpeter, Edwin E.},
  journal =       {\apj},
  month =         jun,
  pages =         {943},
  title =         {The {Interaction} of {X}-{Ray} {Sources} with
                   {Optically} {Thin} {Environments}},
  volume =        {156},
  year =          {1969},
  doi =           {10.1086/150026},
  issn =          {0004-637X},
  url =           {http://adsabs.harvard.edu/abs/1969ApJ...156..943T},
}

@article{temple_high-ionization_2021,
  author =        {Temple, Matthew J. and Ferland, Gary J. and
                   Rankine, Amy L. and Chatzikos, Marios and
                   Hewett, Paul C.},
  journal =       {\mnras},
  month =         aug,
  pages =         {3247--3259},
  title =         {High-ionization emission-line ratios from quasar
                   broad-line regions: metallicity or density?},
  volume =        {505},
  year =          {2021},
  doi =           {10.1093/mnras/stab1610},
  issn =          {0035-8711},
  url =           {https://ui.adsabs.harvard.edu/abs/2021MNRAS.505.3247T},
}

@article{wilkins2025,
       author = {{Wilkins}, Stephen M. and {Vijayan}, Aswin P. and {Hagen}, Scott and {Caruana}, Joseph and {Conselice}, Christopher J. and {Done}, Chris and {Hirschmann}, Michaela and {Irodotou}, Dimitrios and {Lovell}, Christopher C. and {Matthee}, Jorryt and {Plat}, Ad{\`e}le and {Roper}, William J. and {Taylor}, Anthony J.},
        title = "{First Light and Reionization Epoch Simulations (FLARES) -- XVIII: the ionising emissivities and hydrogen recombination line properties of early AGN}",
      journal = {arXiv e-prints},
         year = 2025,
        month = may,
          eid = {arXiv:2505.05257},
        pages = {arXiv:2505.05257},
          doi = {10.48550/arXiv.2505.05257},
archivePrefix = {arXiv},
       eprint = {2505.05257},
 primaryClass = {astro-ph.GA},
       adsurl = {https://ui.adsabs.harvard.edu/abs/2025arXiv250505257W}
}

@article{korista_atlas_1997,
  author =        {Korista, Kirk and Baldwin, Jack and Ferland, Gary and
                   Verner, Dima},
  journal =       {\apjs},
  month =         jan,
  pages =         {401--415},
  title =         {An {Atlas} of {Computed} {Equivalent} {Widths} of
                   {Quasar} {Broad} {Emission} {Lines}},
  volume =        {108},
  year =          {1997},
  doi =           {10.1086/312966},
  url =           {https://ui.adsabs.harvard.edu/abs/1997ApJS..108..401K},
}

@article{emmering_magnetic_1992,
  author =        {Emmering, R. T. and Blandford, R. D. and
                   Shlosman, I.},
  journal =       {\apj},
  month =         feb,
  pages =         {460--477},
  title =         {Magnetic acceleration of broad emission-line clouds
                   in active galactic nuclei},
  volume =        {385},
  year =          {1992},
  doi =           {10.1086/170955},
}

@article{murray_accretion_1995,
  author =        {Murray, N. and Chiang, J. and Grossman, S. A. and
                   Voit, G. M.},
  journal =       {\apj},
  month =         oct,
  pages =         {498},
  title =         {Accretion {Disk} {Winds} from {Active} {Galactic}
                   {Nuclei}},
  volume =        {451},
  year =          {1995},
  doi =           {10.1086/176238},
}

@article{elvis_structure_2000,
  author =        {Elvis, M.},
  journal =       {\apj},
  month =         dec,
  pages =         {63--76},
  title =         {A {Structure} for {Quasars}},
  volume =        {545},
  year =          {2000},
  doi =           {10.1086/317778},
}

@article{sim_multidimensional_2008,
  author =        {Sim, S. A. and Long, K. S. and Miller, L. and
                   Turner, T. J.},
  journal =       {\mnras},
  month =         aug,
  note =          {\_eprint: 0805.2251},
  number =        {2},
  pages =         {611--624},
  title =         {Multidimensional modelling of {X}-ray spectra for
                   {AGN} accretion disc outflows},
  volume =        {388},
  year =          {2008},
  doi =           {10.1111/j.1365-2966.2008.13466.x},
  url =           {https://ui.adsabs.harvard.edu/abs/2008MNRAS.388..611S},
}

@ARTICLE{Choi2022b,
       author = {{Choi}, Hyunseop and {Leighly}, Karen M. and {Dabbieri}, Collin and {Terndrup}, Donald M. and {Gallagher}, Sarah C. and {Richards}, Gordon T.},
        title = "{The Physical Properties of Low-redshift FeLoBAL Quasars. III. The Location and Geometry of the Outflows}",
      journal = {\apj},
         year = 2022,
        month = sep,
       volume = {936},
       number = {2},
          eid = {110},
        pages = {110},
          doi = {10.3847/1538-4357/ac854c},
archivePrefix = {arXiv},
       eprint = {2208.02834},
 primaryClass = {astro-ph.GA},
       adsurl = {https://ui.adsabs.harvard.edu/abs/2022ApJ...936..110C}
}

@ARTICLE{Choi2020,
       author = {{Choi}, Hyunseop and {Leighly}, Karen M. and {Terndrup}, Donald M. and {Gallagher}, Sarah C. and {Richards}, Gordon T.},
        title = "{Discovery of a Remarkably Powerful Broad Absorption-line Quasar Outflow in SDSS J135246.37+423923.5}",
      journal = {\apj},
         year = 2020,
        month = mar,
       volume = {891},
       number = {1},
          eid = {53},
        pages = {53},
          doi = {10.3847/1538-4357/ab6f72},
archivePrefix = {arXiv},
       eprint = {2001.07347},
 primaryClass = {astro-ph.GA},
       adsurl = {https://ui.adsabs.harvard.edu/abs/2020ApJ...891...53C}
}

@ARTICLE{Gaskell2024,
       author = {{Gaskell}, C. Martin and {Gill}, Jake J.~M. and {Singh}, Japneet},
        title = "{Attenuation from the optical to the extreme ultraviolet by dust associated with broad absorption line quasars: The driving force for outflows}",
      journal = {\mnras},
         year = 2024,
        month = sep,
       volume = {533},
       number = {3},
        pages = {3676-3684},
          doi = {10.1093/mnras/stae1886},
       adsurl = {https://ui.adsabs.harvard.edu/abs/2024MNRAS.533.3676G}
}

@ARTICLE{arav2020,
       author = {{Arav}, Nahum and {Xu}, Xinfeng and {Miller}, Timothy and {Kriss}, Gerard A. and {Plesha}, Rachel},
        title = "{HST/COS Observations of Quasar Outflows in the 500-1050 {\r{A}} Rest Frame. I. The Most Energetic Outflows in the Universe and Other Discoveries}",
      journal = {\apjs},
         year = 2020,
        month = apr,
       volume = {247},
       number = {2},
          eid = {37},
        pages = {37},
          doi = {10.3847/1538-4365/ab66af},
archivePrefix = {arXiv},
       eprint = {2003.08688},
 primaryClass = {astro-ph.GA},
       adsurl = {https://ui.adsabs.harvard.edu/abs/2020ApJS..247...37A}
}

@ARTICLE{Dehghanian2025,
       author = {{Dehghanian}, M. and {Arav}, N. and {Sharma}, M. and {Walker}, G. and {Johnston}, K. and {Kaupin}, M.},
        title = "{An energetic absorption outflow in QSO J1402+2330: Analysis of DESI observations}",
      journal = {\aap},
         year = 2025,
        month = mar,
       volume = {695},
          eid = {A4},
        pages = {A4},
          doi = {10.1051/0004-6361/202453384},
archivePrefix = {arXiv},
       eprint = {2501.18034},
 primaryClass = {astro-ph.GA},
       adsurl = {https://ui.adsabs.harvard.edu/abs/2025A&A...695A...4D}
}

@ARTICLE{Sharma2025,
       author = {{Sharma}, Mayank and {Arav}, Nahum and {Korista}, Kirk T. and {Bautista}, Manuel and {Dehghanian}, Maryam and {Byun}, Doyee and {Walker}, Gwen and {Mintz}, Sasha},
        title = "{Physical characterization of the FeLoBAL outflow in SDSS J0932+0840: Analysis of VLT/UVES observations}",
      journal = {\aap},
         year = 2025,
        month = jan,
       volume = {693},
          eid = {A254},
        pages = {A254},
          doi = {10.1051/0004-6361/202452735},
archivePrefix = {arXiv},
       eprint = {2412.06929},
 primaryClass = {astro-ph.GA},
       adsurl = {https://ui.adsabs.harvard.edu/abs/2025A&A...693A.254S}
}

@ARTICLE{arav2018,
       author = {{Arav}, Nahum and {Liu}, Guilin and {Xu}, Xinfeng and {Stidham}, James and {Benn}, Chris and {Chamberlain}, Carter},
        title = "{Evidence that 50\% of BALQSO Outflows Are Situated at Least 100 pc from the Central Source}",
      journal = {\apj},
         year = 2018,
        month = apr,
       volume = {857},
       number = {1},
          eid = {60},
        pages = {60},
          doi = {10.3847/1538-4357/aab494},
archivePrefix = {arXiv},
       eprint = {1805.01543},
 primaryClass = {astro-ph.GA},
       adsurl = {https://ui.adsabs.harvard.edu/abs/2018ApJ...857...60A}
}

@ARTICLE{Chamberlain2015,
       author = {{Chamberlain}, Carter and {Arav}, Nahum},
        title = "{Large-scale outflow in quasar LBQS J1206+1052: HST/COS observations}",
      journal = {\mnras},
         year = 2015,
        month = nov,
       volume = {454},
       number = {1},
        pages = {675-680},
          doi = {10.1093/mnras/stv1979},
       adsurl = {https://ui.adsabs.harvard.edu/abs/2015MNRAS.454..675C}
}

@ARTICLE{Borguet2013,
       author = {{Borguet}, Benoit C.~J. and {Arav}, Nahum and {Edmonds}, Doug and {Chamberlain}, Carter and {Benn}, Chris},
        title = "{Major Contributor to AGN Feedback: VLT X-shooter Observations of S IV BALQSO Outflows}",
      journal = {\apj},
         year = 2013,
        month = jan,
       volume = {762},
       number = {1},
          eid = {49},
        pages = {49},
          doi = {10.1088/0004-637X/762/1/49},
archivePrefix = {arXiv},
       eprint = {1211.6250},
 primaryClass = {astro-ph.CO},
       adsurl = {https://ui.adsabs.harvard.edu/abs/2013ApJ...762...49B}
}

@ARTICLE{drew1982,
       author = {{Drew}, J. and {Giddings}, J.},
        title = "{An evaluation of spherically symmetric wind models for broad absorption line QSOs}",
      journal = {\mnras},
         year = 1982,
        month = oct,
       volume = {201},
        pages = {27-50},
          doi = {10.1093/mnras/201.1.27},
       adsurl = {https://ui.adsabs.harvard.edu/abs/1982MNRAS.201...27D}
}

@ARTICLE{drew1984,
       author = {{Drew}, J.~E. and {Boksenberg}, A.},
        title = "{Optical spectroscopy of two broad absorption line QSOs and implications for spherical symmetric absorbing wind models.}",
      journal = {\mnras},
         year = 1984,
        month = dec,
       volume = {211},
        pages = {813-831},
          doi = {10.1093/mnras/211.4.813},
       adsurl = {https://ui.adsabs.harvard.edu/abs/1984MNRAS.211..813D}
}

@ARTICLE{Mosallanezhad2025,
       author = {{Mosallanezhad}, Amin and {Knigge}, Christian and {Scepi}, Nicolas and {Matthews}, James H. and {Long}, Knox S. and {Sim}, Stuart A. and {Wallis}, Austen},
        title = "{Monte Carlo radiation hydrodynamic simulations of line-driven disc winds: relaxing the isothermal approximation}",
      journal = {\mnras},
         year = 2025,
        month = aug,
       volume = {541},
       number = {3},
        pages = {2393-2404},
          doi = {10.1093/mnras/staf1101},
archivePrefix = {arXiv},
       eprint = {2507.05085},
 primaryClass = {astro-ph.HE},
       adsurl = {https://ui.adsabs.harvard.edu/abs/2025MNRAS.541.2393M}
}

@ARTICLE{Higginbottom2024,
       author = {{Higginbottom}, Nick and {Scepi}, Nicolas and {Knigge}, Christian and {Long}, Knox S. and {Matthews}, James H. and {Sim}, Stuart A.},
        title = "{State-of-the-art simulations of line-driven accretion disc winds: realistic radiation hydrodynamics leads to weaker outflows}",
      journal = {\mnras},
         year = 2024,
        month = jan,
       volume = {527},
       number = {3},
        pages = {9236-9249},
          doi = {10.1093/mnras/stad3830},
archivePrefix = {arXiv},
       eprint = {2312.06042},
 primaryClass = {astro-ph.HE},
       adsurl = {https://ui.adsabs.harvard.edu/abs/2024MNRAS.527.9236H}
}

@ARTICLE{proga2002,
       author = {{Proga}, Daniel and {Kallman}, Timothy R. and {Drew}, Janet E. and {Hartley}, Louise E.},
        title = "{Resonance Line Profile Calculations Based on Hydrodynamical Models of Cataclysmic Variable Winds}",
      journal = {\apj},
         year = 2002,
        month = jun,
       volume = {572},
       number = {1},
        pages = {382-391},
          doi = {10.1086/340339},
archivePrefix = {arXiv},
       eprint = {astro-ph/0202384},
 primaryClass = {astro-ph},
       adsurl = {https://ui.adsabs.harvard.edu/abs/2002ApJ...572..382P}
}

@ARTICLE{proga1998,
       author = {{Proga}, Daniel and {Stone}, James M. and {Drew}, Janet E.},
        title = "{Radiation-driven winds from luminous accretion discs}",
      journal = {\mnras},
         year = 1998,
        month = apr,
       volume = {295},
       number = {3},
        pages = {595-617},
          doi = {10.1046/j.1365-8711.1998.01337.x},
archivePrefix = {arXiv},
       eprint = {astro-ph/9710305},
 primaryClass = {astro-ph},
       adsurl = {https://ui.adsabs.harvard.edu/abs/1998MNRAS.295..595P}
}

@ARTICLE{Nomura2020,
       author = {{Nomura}, Mariko and {Ohsuga}, Ken and {Done}, Chris},
        title = "{Line-driven disc wind in near-Eddington active galactic nuclei: decrease of mass accretion rate due to powerful outflow}",
      journal = {\mnras},
         year = 2020,
        month = may,
       volume = {494},
       number = {3},
        pages = {3616-3626},
          doi = {10.1093/mnras/staa948},
archivePrefix = {arXiv},
       eprint = {1811.01966},
 primaryClass = {astro-ph.HE},
       adsurl = {https://ui.adsabs.harvard.edu/abs/2020MNRAS.494.3616N}
}

@ARTICLE{Dyda2025,
       author = {{Dyda}, Sergei and {Dannen}, Randall C. and {Kallman}, Timothy R. and {Davis}, Shane W. and {Proga}, Daniel},
        title = "{Time-dependent AGN disc winds - II. Effects of photoionization}",
      journal = {\mnras},
         year = 2025,
        month = jul,
       volume = {540},
       number = {3},
        pages = {2612-2622},
          doi = {10.1093/mnras/staf898},
archivePrefix = {arXiv},
       eprint = {2504.00117},
 primaryClass = {astro-ph.HE},
       adsurl = {https://ui.adsabs.harvard.edu/abs/2025MNRAS.540.2612D}
}

@ARTICLE{dyda2024,
       author = {{Dyda}, Sergei and {Davis}, Shane W. and {Proga}, Daniel},
        title = "{Time-dependent AGN disc winds - I. X-ray irradiation}",
      journal = {\mnras},
         year = 2024,
        month = jun,
       volume = {530},
       number = {4},
        pages = {5143-5154},
          doi = {10.1093/mnras/stae1159},
archivePrefix = {arXiv},
       eprint = {2310.18557},
 primaryClass = {astro-ph.HE},
       adsurl = {https://ui.adsabs.harvard.edu/abs/2024MNRAS.530.5143D}
}

@ARTICLE{Choi2022,
       author = {{Choi}, Hyunseop and {Leighly}, Karen M. and {Terndrup}, Donald M. and {Dabbieri}, Collin and {Gallagher}, Sarah C. and {Richards}, Gordon T.},
        title = "{The Physical Properties of Low-redshift FeLoBAL Quasars. I. Spectral-synthesis Analysis of the Broad Absorption-line (BAL) Outflows Using SimBAL}",
      journal = {\apj},
         year = 2022,
        month = oct,
       volume = {937},
       number = {2},
          eid = {74},
        pages = {74},
          doi = {10.3847/1538-4357/ac61d9},
archivePrefix = {arXiv},
       eprint = {2203.11964},
 primaryClass = {astro-ph.GA},
       adsurl = {https://ui.adsabs.harvard.edu/abs/2022ApJ...937...74C}
}

@ARTICLE{Fabian1999,
       author = {{Fabian}, A.~C.},
        title = "{The obscured growth of massive black holes}",
      journal = {\mnras},
         year = 1999,
        month = oct,
       volume = {308},
       number = {4},
        pages = {L39-L43},
          doi = {10.1046/j.1365-8711.1999.03017.x},
archivePrefix = {arXiv},
       eprint = {astro-ph/9908064},
 primaryClass = {astro-ph},
       adsurl = {https://ui.adsabs.harvard.edu/abs/1999MNRAS.308L..39F}
}

@ARTICLE{Springel2005,
       author = {{Springel}, Volker and {Di Matteo}, Tiziana and {Hernquist}, Lars},
        title = "{Modelling feedback from stars and black holes in galaxy mergers}",
      journal = {\mnras},
         year = 2005,
        month = aug,
       volume = {361},
       number = {3},
        pages = {776-794},
          doi = {10.1111/j.1365-2966.2005.09238.x},
archivePrefix = {arXiv},
       eprint = {astro-ph/0411108},
 primaryClass = {astro-ph},
       adsurl = {https://ui.adsabs.harvard.edu/abs/2005MNRAS.361..776S}
}

@ARTICLE{Sijacki2007,
       author = {{Sijacki}, Debora and {Springel}, Volker and {Di Matteo}, Tiziana and {Hernquist}, Lars},
        title = "{A unified model for AGN feedback in cosmological simulations of structure formation}",
      journal = {\mnras},
         year = 2007,
        month = sep,
       volume = {380},
       number = {3},
        pages = {877-900},
          doi = {10.1111/j.1365-2966.2007.12153.x},
archivePrefix = {arXiv},
       eprint = {0705.2238},
 primaryClass = {astro-ph},
       adsurl = {https://ui.adsabs.harvard.edu/abs/2007MNRAS.380..877S}
}

@ARTICLE{Sijacki2015,
       author = {{Sijacki}, Debora and {Vogelsberger}, Mark and {Genel}, Shy and {Springel}, Volker and {Torrey}, Paul and {Snyder}, Gregory F. and {Nelson}, Dylan and {Hernquist}, Lars},
        title = "{The Illustris simulation: the evolving population of black holes across cosmic time}",
      journal = {\mnras},
         year = 2015,
        month = sep,
       volume = {452},
       number = {1},
        pages = {575-596},
          doi = {10.1093/mnras/stv1340},
archivePrefix = {arXiv},
       eprint = {1408.6842},
 primaryClass = {astro-ph.GA},
       adsurl = {https://ui.adsabs.harvard.edu/abs/2015MNRAS.452..575S}
}

@article{sim_multidimensional_2010,
  author =        {Sim, S. A. and Proga, D. and Miller, L. and
                   Long, K. S. and Turner, T. J.},
  journal =       {\mnras},
  month =         nov,
  note =          {\_eprint: 1006.3449},
  number =        {3},
  pages =         {1396--1408},
  title =         {Multidimensional modelling of {X}-ray spectra for
                   {AGN} accretion disc outflows - {III}. {Application}
                   to a hydrodynamical simulation},
  volume =        {408},
  year =          {2010},
  doi =           {10.1111/j.1365-2966.2010.17215.x},
  issn =          {0035-8711},
  url =           {https://ui.adsabs.harvard.edu/abs/2010MNRAS.408.1396S},
}

@article{hagino_origin_2015,
  author =        {Hagino, Kouichi and Odaka, Hirokazu and Done, Chris and
                   Gandhi, Poshak and Watanabe, Shin and Sako, Masao and
                   Takahashi, Tadayuki},
  journal =       {\mnras},
  month =         jan,
  note =          {\_eprint: 1410.1640},
  number =        {1},
  pages =         {663--676},
  title =         {The origin of ultrafast outflows in {AGN}: {Monte}
                   {Carlo} simulations of the wind in {PDS} 456},
  volume =        {446},
  year =          {2015},
  doi =           {10.1093/mnras/stu2095},
  language =      {en},
  url =           {https://ui.adsabs.harvard.edu/abs/2015MNRAS.446..663H/
                  abstract},
}

@article{shlosman_winds_1993,
  author =        {Shlosman, I. and Vitello, P.},
  journal =       {\apj},
  month =         may,
  pages =         {372--386},
  title =         {Winds from accretion disks - {Ultraviolet} line
                   formation in cataclysmic variables},
  volume =        {409},
  year =          {1993},
  doi =           {10.1086/172670},
}

@article{higginbottom_simple_2013,
  author =        {Higginbottom, N. and Knigge, C. and Long, K. S. and
                   Sim, S. A. and Matthews, J. H.},
  journal =       {\mnras},
  month =         dec,
  pages =         {1390--1407},
  title =         {A simple disc wind model for broad absorption line
                   quasars},
  volume =        {436},
  year =          {2013},
  doi =           {10.1093/mnras/stt1658},
  issn =          {0035-8711},
  url =           {https://ui.adsabs.harvard.edu/abs/2013MNRAS.436.1390H},
}

@ARTICLE{lucy2003,
       author = {{Lucy}, L.~B.},
        title = "{Monte Carlo transition probabilities. II.}",
      journal = {\aap},
         year = 2003,
        month = may,
       volume = {403},
        pages = {261-275},
          doi = {10.1051/0004-6361:20030357},
archivePrefix = {arXiv},
       eprint = {astro-ph/0303202},
 primaryClass = {astro-ph},
       adsurl = {https://ui.adsabs.harvard.edu/abs/2003A&A...403..261L}
}

@article{yong_black_2016,
  author =        {Yong, Suk Yee and Webster, Rachel L. and
                   King, Anthea L.},
  journal =       {Publications of the Astronomical Society of
                   Australia},
  month =         mar,
  pages =         {e009},
  title =         {Black {Hole} {Mass} {Estimation}: {How} {Good} is the
                   {Virial} {Estimate}?},
  volume =        {33},
  year =          {2016},
  doi =           {10.1017/pasa.2016.8},
  issn =          {1323-3580},
  url =           {http://adsabs.harvard.edu/abs/2016PASA...33....9Y},
}

@article{matthews_testing_2016,
  author =        {Matthews, J. H. and Knigge, C. and Long, K. S. and
                   Sim, S. A. and Higginbottom, N. and Mangham, S. W.},
  journal =       {\mnras},
  month =         may,
  pages =         {293--305},
  title =         {Testing quasar unification: radiative transfer in
                   clumpy winds},
  volume =        {458},
  year =          {2016},
  doi =           {10.1093/mnras/stw323},
  issn =          {0035-8711},
  url =           {https://ui.adsabs.harvard.edu/abs/2016MNRAS.458..293M},
}

@ARTICLE{Baldwin1996,
       author = {{Baldwin}, J.~A. and {Ferland}, G.~J. and {Korista}, K.~T. and {Carswell}, R.~F. and {Hamann}, F. and {Phillips}, M.~M. and {Verner}, D. and {Wilkes}, Belinda J. and {Williams}, R.~E.},
        title = "{Very High Density Clumps and Outflowing Winds in QSO Broad-Line Regions}",
      journal = {\apj},
         year = 1996,
        month = apr,
       volume = {461},
        pages = {664},
          doi = {10.1086/177093},
       adsurl = {https://ui.adsabs.harvard.edu/abs/1996ApJ...461..664B}
}

@ARTICLE{Leighly2004b,
       author = {{Leighly}, Karen M.},
        title = "{Hubble Space Telescope STIS Ultraviolet Spectral Evidence of Outflow in Extreme Narrow-Line Seyfert 1 Galaxies. II. Modeling and Interpretation}",
      journal = {\apj},
         year = 2004,
        month = aug,
       volume = {611},
       number = {1},
        pages = {125-152},
          doi = {10.1086/422089},
archivePrefix = {arXiv},
       eprint = {astro-ph/0402452},
 primaryClass = {astro-ph},
       adsurl = {https://ui.adsabs.harvard.edu/abs/2004ApJ...611..125L}
}

@ARTICLE{Leighly2004a,
       author = {{Leighly}, Karen M. and {Moore}, John R.},
        title = "{Hubble Space Telescope STIS Ultraviolet Spectral Evidence of Outflow in Extreme Narrow-Line Seyfert 1 Galaxies. I. Data and Analysis}",
      journal = {\apj},
         year = 2004,
        month = aug,
       volume = {611},
       number = {1},
        pages = {107-124},
          doi = {10.1086/422088},
archivePrefix = {arXiv},
       eprint = {astro-ph/0402453},
 primaryClass = {astro-ph},
       adsurl = {https://ui.adsabs.harvard.edu/abs/2004ApJ...611..107L}
}

@ARTICLE{Rankine2026,
       author = {{Rankine}, Amy L. and {Homan}, David and {Aird}, James and {Hiremath}, Pranavi and {Anderson}, Scott F. and {Assef}, Roberto J. and {Bauer}, Franz E. and {Brandt}, W.~N. and {Brusa}, Marcella and {Buchner}, Johannes and {Chira}, Maria and {D{\'\i}az}, Yaherlyn and {Hall}, Patrick B. and {Koekemoer}, Anton M. and {Krumpe}, Mirko and {Lamer}, Georg and {Liu}, Teng and {Morrison}, Sean and {Musiimenta}, Blessing and {Negrete}, C.~A. and {Ni}, Qingling and {Rodr{\'\i}guez Hidalgo}, Paola and {Salvato}, Mara and {Schneider}, Donald P. and {Shen}, Yue and {Temple}, Matthew J. and {Tub{\'\i}n-Arenas}, Dus{\'a}n and {Wylezalek}, Dominika},
        title = "{C IV wind properties of the SDSS-V X-ray selected quasars: strong optical-to-UV emission is key regardless of X-ray strength}",
      journal = {\mnras},
         year = 2026,
        month = jun,
       volume = {548},
       number = {4},
          eid = {stag734},
        pages = {stag734},
          doi = {10.1093/mnras/stag734},
archivePrefix = {arXiv},
       eprint = {2603.15075},
 primaryClass = {astro-ph.GA},
       adsurl = {https://ui.adsabs.harvard.edu/abs/2026MNRAS.548ag734R}
}

@ARTICLE{Hiremath2025,
       author = {{Hiremath}, Pranavi and {Rankine}, Amy L. and {Aird}, James and {Brandt}, W.~N. and {Rodr{\'\i}guez Hidalgo}, Paola and {Anderson}, Scott F. and {Aydar}, Catarina and {Ricci}, Claudio and {Schneider}, Donald P. and {Vivek}, M. and {Igo}, Zsofi and {Morrison}, Sean and {Salvato}, Mara},
        title = "{X-ray selected broad absorption line quasars in SDSS-V: BALs and non-BALs span the same range of X-ray properties}",
      journal = {\mnras},
         year = 2025,
        month = sep,
       volume = {542},
       number = {3},
        pages = {2105-2127},
          doi = {10.1093/mnras/staf1352},
archivePrefix = {arXiv},
       eprint = {2508.13682},
 primaryClass = {astro-ph.GA},
       adsurl = {https://ui.adsabs.harvard.edu/abs/2025MNRAS.542.2105H}
}

@ARTICLE{Shlentsova2026,
       author = {{Shlentsova}, Anastasia and {Trefoloni}, Bartolomeo and {Signorini}, Matilde and {Risaliti}, Guido and {Lusso}, Elisabeta and {Nardini}, Emanuele and {Bauer}, Franz E. and {Temple}, Matthew J. and {Rankine}, Amy L. and {Richards}, Gordon T.},
        title = "{The X-ray properties of the most luminous quasars with strong emission-line outflows}",
      journal = {\aap},
         year = 2026,
        month = mar,
       volume = {707},
          eid = {A313},
        pages = {A313},
          doi = {10.1051/0004-6361/202555381},
archivePrefix = {arXiv},
       eprint = {2602.06793},
 primaryClass = {astro-ph.GA},
       adsurl = {https://ui.adsabs.harvard.edu/abs/2026A&A...707A.313S}
}

@ARTICLE{Chakravorty2016,
       author = {{Chakravorty}, S. and {Petrucci}, P.-O. and {Ferreira}, J. and {Henri}, G. and {Belmont}, R. and {Clavel}, M. and {Corbel}, S. and {Rodriguez}, J. and {Coriat}, M. and {Drappeau}, S. and {Malzac}, J.},
        title = "{Absorption lines from magnetically driven winds in X-ray binaries}",
      journal = {\aap},
         year = 2016,
        month = may,
       volume = {589},
          eid = {A119},
        pages = {A119},
          doi = {10.1051/0004-6361/201527163},
archivePrefix = {arXiv},
       eprint = {1512.09149},
 primaryClass = {astro-ph.HE},
       adsurl = {https://ui.adsabs.harvard.edu/abs/2016A&A...589A.119C}
}

@article{hunter07,
	adsurl = {https://ui.adsabs.harvard.edu/abs/2007CSE.....9...90H},
	author = {{Hunter}, John D.},
	doi = {10.1109/MCSE.2007.55},
	journal = {Computing in Science and Engineering},
	month = may,
	number = {3},
	pages = {90-95},
	title = {{Matplotlib: A 2D Graphics Environment}},
	volume = {9},
	year = 2007}

@article{ferland17,
	adsurl = {https://ui.adsabs.harvard.edu/abs/2017RMxAA..53..385F},
	archiveprefix = {arXiv},
	author = {{Ferland}, G.~J. and {Chatzikos}, M. and {Guzm{\'a}n}, F. and {Lykins}, M.~L. and {van Hoof}, P.~A.~M. and {Williams}, R.~J.~R. and {Abel}, N.~P. and {Badnell}, N.~R. and {Keenan}, F.~P. and {Porter}, R.~L. and {Stancil}, P.~C.},
	eprint = {1705.10877},
	journal = {\rmxaa},
	month = oct,
	pages = {385-438},
	primaryclass = {astro-ph.GA},
	title = {{The 2017 Release Cloudy}},
	volume = {53},
	year = 2017}

@article{astropy-collaboration13,
	adsurl = {https://ui.adsabs.harvard.edu/abs/2013A&A...558A..33A},
	archiveprefix = {arXiv},
	author = {{Astropy Collaboration} and {Robitaille}, Thomas P. and {Tollerud}, Erik J. and {Greenfield}, Perry and {Droettboom}, Michael and {Bray}, Erik and {Aldcroft}, Tom and {Davis}, Matt and {Ginsburg}, Adam and {Price-Whelan}, Adrian M. and {Kerzendorf}, Wolfgang E. and {Conley}, Alexander and {Crighton}, Neil and {Barbary}, Kyle and {Muna}, Demitri and {Ferguson}, Henry and {Grollier}, Fr{\'e}d{\'e}ric and {Parikh}, Madhura M. and {Nair}, Prasanth H. and {Unther}, Hans M. and {Deil}, Christoph and {Woillez}, Julien and {Conseil}, Simon and {Kramer}, Roban and {Turner}, James E.~H. and {Singer}, Leo and {Fox}, Ryan and {Weaver}, Benjamin A. and {Zabalza}, Victor and {Edwards}, Zachary I. and {Azalee Bostroem}, K. and {Burke}, D.~J. and {Casey}, Andrew R. and {Crawford}, Steven M. and {Dencheva}, Nadia and {Ely}, Justin and {Jenness}, Tim and {Labrie}, Kathleen and {Lim}, Pey Lian and {Pierfederici}, Francesco and {Pontzen}, Andrew and {Ptak}, Andy and {Refsdal}, Brian and {Servillat}, Mathieu and {Streicher}, Ole},
	doi = {10.1051/0004-6361/201322068},
	eid = {A33},
	eprint = {1307.6212},
	journal = {\aap},
	month = oct,
	pages = {A33},
	primaryclass = {astro-ph.IM},
	title = {{Astropy: A community Python package for astronomy}},
	volume = {558},
	year = 2013}

@article{astropy-collaboration18,
	adsurl = {https://ui.adsabs.harvard.edu/abs/2018AJ....156..123A},
	archiveprefix = {arXiv},
	author = {{Astropy Collaboration} and {Price-Whelan}, A.~M. and {Sip{\H{o}}cz}, B.~M. and {G{\"u}nther}, H.~M. and {Lim}, P.~L. and {Crawford}, S.~M. and {Conseil}, S. and {Shupe}, D.~L. and {Craig}, M.~W. and {Dencheva}, N. and {Ginsburg}, A. and {Vand erPlas}, J.~T. and {Bradley}, L.~D. and {P{\'e}rez-Su{\'a}rez}, D. and {de Val-Borro}, M. and {Aldcroft}, T.~L. and {Cruz}, K.~L. and {Robitaille}, T.~P. and {Tollerud}, E.~J. and {Ardelean}, C. and {Babej}, T. and {Bach}, Y.~P. and {Bachetti}, M. and {Bakanov}, A.~V. and {Bamford}, S.~P. and {Barentsen}, G. and {Barmby}, P. and {Baumbach}, A. and {Berry}, K.~L. and {Biscani}, F. and {Boquien}, M. and {Bostroem}, K.~A. and {Bouma}, L.~G. and {Brammer}, G.~B. and {Bray}, E.~M. and {Breytenbach}, H. and {Buddelmeijer}, H. and {Burke}, D.~J. and {Calderone}, G. and {Cano Rodr{\'\i}guez}, J.~L. and {Cara}, M. and {Cardoso}, J.~V.~M. and {Cheedella}, S. and {Copin}, Y. and {Corrales}, L. and {Crichton}, D. and {D'Avella}, D. and {Deil}, C. and {Depagne}, {\'E}. and {Dietrich}, J.~P. and {Donath}, A. and {Droettboom}, M. and {Earl}, N. and {Erben}, T. and {Fabbro}, S. and {Ferreira}, L.~A. and {Finethy}, T. and {Fox}, R.~T. and {Garrison}, L.~H. and {Gibbons}, S.~L.~J. and {Goldstein}, D.~A. and {Gommers}, R. and {Greco}, J.~P. and {Greenfield}, P. and {Groener}, A.~M. and {Grollier}, F. and {Hagen}, A. and {Hirst}, P. and {Homeier}, D. and {Horton}, A.~J. and {Hosseinzadeh}, G. and {Hu}, L. and {Hunkeler}, J.~S. and {Ivezi{\'c}}, {\v{Z}}. and {Jain}, A. and {Jenness}, T. and {Kanarek}, G. and {Kendrew}, S. and {Kern}, N.~S. and {Kerzendorf}, W.~E. and {Khvalko}, A. and {King}, J. and {Kirkby}, D. and {Kulkarni}, A.~M. and {Kumar}, A. and {Lee}, A. and {Lenz}, D. and {Littlefair}, S.~P. and {Ma}, Z. and {Macleod}, D.~M. and {Mastropietro}, M. and {McCully}, C. and {Montagnac}, S. and {Morris}, B.~M. and {Mueller}, M. and {Mumford}, S.~J. and {Muna}, D. and {Murphy}, N.~A. and {Nelson}, S. and {Nguyen}, G.~H. and {Ninan}, J.~P. and {N{\"o}the}, M. and {Ogaz}, S. and {Oh}, S. and {Parejko}, J.~K. and {Parley}, N. and {Pascual}, S. and {Patil}, R. and {Patil}, A.~A. and {Plunkett}, A.~L. and {Prochaska}, J.~X. and {Rastogi}, T. and {Reddy Janga}, V. and {Sabater}, J. and {Sakurikar}, P. and {Seifert}, M. and {Sherbert}, L.~E. and {Sherwood-Taylor}, H. and {Shih}, A.~Y. and {Sick}, J. and {Silbiger}, M.~T. and {Singanamalla}, S. and {Singer}, L.~P. and {Sladen}, P.~H. and {Sooley}, K.~A. and {Sornarajah}, S. and {Streicher}, O. and {Teuben}, P. and {Thomas}, S.~W. and {Tremblay}, G.~R. and {Turner}, J.~E.~H. and {Terr{\'o}n}, V. and {van Kerkwijk}, M.~H. and {de la Vega}, A. and {Watkins}, L.~L. and {Weaver}, B.~A. and {Whitmore}, J.~B. and {Woillez}, J. and {Zabalza}, V. and {Astropy Contributors}},
	doi = {10.3847/1538-3881/aabc4f},
	eid = {123},
	eprint = {1801.02634},
	journal = {\aj},
	month = sep,
	number = {3},
	pages = {123},
	primaryclass = {astro-ph.IM},
	title = {{The Astropy Project: Building an Open-science Project and Status of the v2.0 Core Package}},
	volume = {156},
	year = 2018}

@misc{Galassi2018_gsl,
	author = {Galassi, M. et al},
	title = {GNU Scientific Library Reference Manual},
	url = {https://www.gnu.org/software/gsl/},
	year = 2018}

@ARTICLE{gallagher2005,
       author = {{Gallagher}, S.~C. and {Richards}, Gordon T. and {Hall}, Patrick B. and {Brandt}, W.~N. and {Schneider}, Donald P. and {Vanden Berk}, Daniel E.},
        title = "{X-Ray Insights into Interpreting C IV Blueshifts and Optical/Ultraviolet Continua}",
      journal = {\aj},
         year = 2005,
        month = feb,
       volume = {129},
       number = {2},
        pages = {567-577},
          doi = {10.1086/426913},
archivePrefix = {arXiv},
       eprint = {astro-ph/0410641},
 primaryClass = {astro-ph},
       adsurl = {https://ui.adsabs.harvard.edu/abs/2005AJ....129..567G}
}

@article{long_modeling_2002,
  author =        {Long, K. S. and Knigge, C.},
  journal =       {\apj},
  month =         nov,
  pages =         {725--740},
  title =         {Modeling the {Spectral} {Signatures} of {Accretion}
                   {Disk} {Winds}: {A} {New} {Monte} {Carlo} {Approach}},
  volume =        {579},
  year =          {2002},
  doi =           {10.1086/342879},
}

@article{sirocco,
  author =        {{Matthews}, James H. and {Long}, Knox S. and
                   {Knigge}, Christian and {Sim}, Stuart A. and
                   {Parkinson}, Edward J. and {Higginbottom}, Nick and
                   {Mangham}, Samuel W. and {Scepi}, Nicolas and
                   {Wallis}, Austen and {Hewitt}, Henrietta A. and
                   {Mosallanezhad}, Amin},
  journal =       {\mnras},
  month =         jan,
  number =        {1},
  pages =         {879-904},
  title =         {{SIROCCO: a publicly available Monte Carlo ionization
                   and radiative transfer code for astrophysical
                   outflows}},
  volume =        {536},
  year =          {2025},
  doi =           {10.1093/mnras/stae2677},
}

@article{coatman2016,
  author =        {{Coatman}, Liam and {Hewett}, Paul C. and
                   {Banerji}, Manda and {Richards}, Gordon T.},
  journal =       {\mnras},
  month =         sep,
  number =        {1},
  pages =         {647-665},
  title =         {{C IV emission-line properties and systematic trends
                   in quasar black hole mass estimates}},
  volume =        {461},
  year =          {2016},
  doi =           {10.1093/mnras/stw1360},
}

@article{chajet_magnetohydrodynamic_2013,
  author =        {Chajet, L. S. and Hall, P. B.},
  journal =       {\mnras},
  month =         mar,
  pages =         {3214--3229},
  title =         {Magnetohydrodynamic disc winds and linewidth
                   distributions},
  volume =        {429},
  year =          {2013},
  doi =           {10.1093/mnras/sts580},
  url =           {https://ui.adsabs.harvard.edu/abs/2013MNRAS.429.3214C},
}

@article{chajet_magnetohydrodynamic_2017,
  author =        {Chajet, L. S. and Hall, P. B.},
  journal =       {\mnras},
  month =         feb,
  pages =         {1741--1756},
  title =         {Magnetohydrodynamic disc winds and line width
                   distributions - {II}},
  volume =        {465},
  year =          {2017},
  doi =           {10.1093/mnras/stw2626},
  url =           {https://ui.adsabs.harvard.edu/abs/2017MNRAS.465.1741C},
}

@inproceedings{proga_theory_2005,
  author =        {Proga, D.},
  booktitle =     {The {Astrophysics} of {Cataclysmic} {Variables} and
                   {Related} {Objects}},
  editor =        {Hameury, J.-M. and Lasota, J.-P.},
  month =         aug,
  pages =         {103},
  series =        {Astronomical {Society} of the {Pacific} {Conference}
                   {Series}},
  title =         {Theory of {Outflows} in {Cataclysmic} {Variables}},
  volume =        {330},
  year =          {2005},
}

@article{shen2013,
  author =        {{Shen}, Yue},
  journal =       {Bulletin of the Astronomical Society of India},
  month =         mar,
  number =        {1},
  pages =         {61-115},
  title =         {{The mass of quasars}},
  volume =        {41},
  year =          {2013},
  doi =           {10.48550/arXiv.1302.2643},
}


\bsp	
\label{lastpage}
\end{document}